\documentclass[printer]{aa}

\bibpunct{(}{)}{;}{a}{}{,}

\usepackage{graphicx}
\usepackage{txfonts}
\usepackage{soul}
\usepackage{txfonts}
\usepackage{color}
\usepackage{ulem}

\def\ks{km s$^{-1}$}

\def\s{$^{\prime\prime}$}

\def\cm3{cm$^{-3}$}

\def\2{$^{12}$CO}
\def\3{$^{13}$CO}
\def\8{C$^{18}$O}

\def\cm2{cm$^{-2}$}

\begin{document}

\title{Cyano radical emission at small spatial scales towards massive protostars}

\author {S. Paron \inst{1}
\and M. E. Ortega \inst{1}
\and A. Marinelli \inst{1}
\and M. B. Areal \inst{1}
\and N. C. Martinez \inst{2}
}

\institute{CONICET - Universidad de Buenos Aires. Instituto de Astronom\'{\i}a y F\'{\i}sica del Espacio
             CC 67, Suc. 28, 1428 Buenos Aires, Argentina
             \email{sparon@iafe.uba.ar}
\and Universidad de Buenos Aires, Facultad de Ciencias Exactas y Naturales, Buenos Aires, Argentina
}

\offprints{S. Paron}

   \date{Received <date>; Accepted <date>}

\abstract
{The cyano radical (CN), one of the first detected interstellar molecular species, is a key 
molecule in many astrochemical chains. Particularly, it is detected towards molecular cores, 
the birth places of the stars, and it is known that it is involved in the rich chemistry that takes
place in these sites.} {At present there are not so many studies about the emission of this molecular species at small spatial scales towards massive young stellar objects. Thus, we present a high-angular resolution CN study towards a sample of massive protostars, with the aim of unveiling the spatial distribution at the small scale of the emission of this radical in relation to the star-forming processes. 
}
{The interstellar CN has a strong emission line at the rest frequency 226874.764 MHz, thus, we search for observing
projects in the ALMA database regarding high-mass star-forming regions observed at Band 6. The used data set was 
observed in the ALMA Cycle 3 with angular and spectral resolutions 
of 0\farcs7 and 1.13 MHz, respectively. A sample of ten high-mass star-forming regions located at the first Galactic quadrant were selected in base on that they present a clear emission of CN at the mentioned frequency.
} 
{We found that the CN traces both molecular condensations and diffuse and extended gas surrounding them. In general, the molecular condensations traced by the maximums of the CN emission do not spatially coincide with the peaks of the continuum emission at 1.3 mm, which trace the molecular cores where the massive stars born.
Based on the presence or lack of near-IR emission associated with such cores, we suggest that our sample is composed by sources at different stages of evolution. The CN is present at both, suggesting that this radical 
may be ubiquitous along the different star formation stages, and hence it may be involved in different chemical reactions occurring along the time in the formation of the stars. Additionally, other molecules such as CH$_{3}$OCHO and CH$_{2}$CHCN were detected towards the continuum peaks of some of the analyzed cores. We found that the missing flux coming from  extended spatial-scales that are filtered out by the interferometer is an important issue to take into account in the analysis of some spectral features and the spatial distribution of the emission.
}{}

\titlerunning{CN emission towards massive protostars}
\authorrunning{S. Paron et al.}


\maketitle
%

\section{Introduction}

Chemical studies towards protostars are important because they 
provide an useful tool for investigating the physical processes involved in the formation of massive and low-mass stars (see \citealt{gerner14} and \citealt{jor04}, respectively). In particular, the evolution of the high-mass star-forming regions is accompanied by an increasing in the chemical complexity in the molecular environment (e.g. \citealt{tan14,van98}). Thus, high-mass star-forming regions are a very suitable laboratory to study astrochemistry, and particularly, the formation of complex molecules \citep{coletta20}.

The cyano radical (CN), one of the first detected interstellar molecular species \citep{adams41,mckellar40}, is a key molecule in many astrochemical chains. For instance, it is related to the chemistry of HCN and HNC, which are ubiquitous species in different interstellar environments \citep{loison14}. Given that CN is very reactive with molecules possessing C=C double and C$\equiv$C triple bonds, it is involved in the formation of cyanopolyynes \citep{gans17}, molecules observed in high-mass star-forming regions and hot cores, and proposed as `chemical clocks' in determining the age of this kind of sources \citep{tani18,chap09}.

From interferometric observations, the CN emission has been used to study protoplaneatry disks \citep{villa19,vanter19,oberg11}
and some hot molecular cores \citep{qiu12,zapata08}. \citet{beuther04}, based on the investigation towards two massive star-forming
regions, pointed out that CN does not trace the central molecular condensations, but mainly gas in their
near vicinity. They suggested that CN appears not well suited for disk studies in massive protostars as in the case of the low-mass stars, and thus, they recommended the use of different molecules. Additionally, it was proposed that CN probes material in the boundary
between the bulk protostellar envelope and its outflow cavity \citep{jor04}. Besides some of the mentioned works, in the literature there are not abundant works regarding the detection of CN towards massive protostars or high-mass star-forming regions using interferometric observations. \citet{han15} presented a study of CN, HCN and HNC towards 38 high-mass star-forming regions, which included high-mass starless cores (HMSC), high-mass protostellar objects (HMPO), and HII regions, but using single-dish observations. Hence, we point out that it is necessary studying the CN emission with high-angular resolution towards a sample of massive protostars with the aim of unveiling the spatial distribution at  the small scale of the emission of this molecular species in relation to the star-forming processes. The Atacama Large Millimeter Array (ALMA) database offers the opportunity to perform that.    

According to the NIST database\footnote{https://dx.doi.org/10.18434/T4JP4Q}, the interstellar CN has a strong emission line at the rest frequency 226874.764 MHz, corresponding to the N$=$2--1 J$=$5/2--3/2 F$=$7/2--5/2 transition, which 
can be blended with the F$=$5/2--3/2 and F$=$3/2--1/2 lines (at 226874.183 and 226875.896 MHz, respectively). 
Thus, we search for observing
projects in the ALMA database regarding high-mass star-forming regions observed at Band 6. In the following section we describe the data used in this work and the source selection.

\section{Data and source selection}
\label{dataS}

Data were obtained from the ALMA Science Archive\footnote{http://almascience.eso.org/aq/}. We used data from the project entitled
`Tracing the evolution of massive protostars' (Code: 2015.1.01312.S, PI: G. Fuller), that was observed in the ALMA Cycle 3 in configurations C36-2 and C36-3 in the 12~m array, at Band 6. We used the calibrated data which 
passed the second level of Quality Assurance (QA2). The theoretical maximum recoverable scale is about 6~arcsec. The angular resolution of this data set is about 0\farcs7, with a beam almost circular of 0\farcs6$\times$0\farcs8 (PA=82\fdg9), and the spectral resolution is 1.13 MHz.  
The Common Astronomy Software Applications (CASA) was used to handle and analyse the data. The task {\it imcontsub} was used to subtract the continuum from the spectral lines and a first order polynomial was used.
The 1$\sigma$ rms noise level is about 1.5 mJy beam$^{-1}$ for both the continuum and the line emission. 

It is important to remark that even though the data passed the QA2 quality level, which assures a reliable calibration for a ``science ready'' data, the automatic pipeline imaging process may give raise to a clean image with some problems/artefacts. For example, an inappropriate setting of the parameters of the {\it clean} task in CASA could generate artificial dips in the spectra. Thus, we reprocessed the raw data set of sources whose spectra exhibit conspicuous dips in, or close to, the main component of the CN emission to check whether the imaging process is responsible of such spectral features. Particular care was taken with the {\it clean} task. The images obtained from our data reprocessing are very similar to those obtained from the archival, and hence
we conclude that such absorption features are not due to the {\it clean}.

The ALMA programme 2015.1.01312.S observed several young high-mass embedded sources based on previous
surveys of the Galactic plane that have identified large populations of candidate young massive protostars. 
From these sources, we selected a sample that have been observed in the spectral window in which the above 
mentioned CN line lies. Additionally, we selected sources whose spectra present
conspicuous lines that allow us to properly identify the CN emission. The sources, with their velocities, distances, and the
spatial resolution of the data in each case are presented in Table\,\ref{tabsources}. The velocity and distance were obtained from \citet{wie12}, who observed 
ammonia towards cold high-mass clumps in the inner Galactic disk. Given that G23.389 and G34.821 are not included
in the Wienen et al.'s survey, the corresponding velocity and distance were obtained from \citet{maud15} and \citet{navarete19}, respectively. 

Additionally, we used near-IR data at the {\it Ks}-band obtained from the the UKIRT InfraRed Deep Sky Surveys (UKIDSS)\footnote{http://wsa.roe.ac.uk/index.html}, release DR11,  and IRAC-{\it Spitzer} data  
obtained from the Galactic Legacy Infrared Mid-Plane Survey Extraordinaire\footnote{https://irsa.ipac.caltech.edu/data/SPITZER/GLIMPSE/}.

\begin{table}
\centering
\tiny
\caption{Sample of studied sources}
\label{tabsources}
\begin{tabular}{lccccc}

\hline\hline
Source      & RA         & Dec.          & v$_{LSR}$ & Distance & Resol.         \\
            & (J2000)    & (J2000)       &  (\ks)    &  (kpc)   &   (pc)          \\
\hline                 
G13.656     & 18:17:24.2 & $-$17:22:12.6 &  47.0     &  4.1     &  0.014       \\
G18.825     & 18:26:58.8 & $-$12:44:43.9 &  62.9     &  4.3     &  0.014       \\
G20.747     & 18:29:16.3 & $-$10:52:10.7 &  59.0     &  3.9     &  0.013        \\
G20.762     & 18:29:12.1 & $-$10:50:34.8 &  57.0     &  3.9     &  0.013        \\
G23.389     & 18:33:14.3 & $-$08:23:57.6 &  75.4     &  4.5     &  0.015         \\
G24.464     & 18:35:11.2 & $-$07:26:31.7 &  119.0    &  6.2     &  0.021          \\
G29.862     & 18:45:59.5 & $-$02:45:06.8 &  100.0    &  5.6     &  0.019          \\
G30.198     & 18:47:03.0 & $-$03:30:35.9 &  105.9    &  5.9     &  0.020          \\
G33.133     & 18:52:08:1 & $+$00:08:12.9 &  76.0     &  4.5     &  0.015           \\
G34.821     & 18:53:37.9 & $+$01:50:28.7 &  56.5     &  1.6     &  0.005           \\
\hline
\end{tabular}
\end{table}

\section{Results}
\label{results}

\begin{figure}[h!]
    \centering
    \includegraphics[width=9cm]{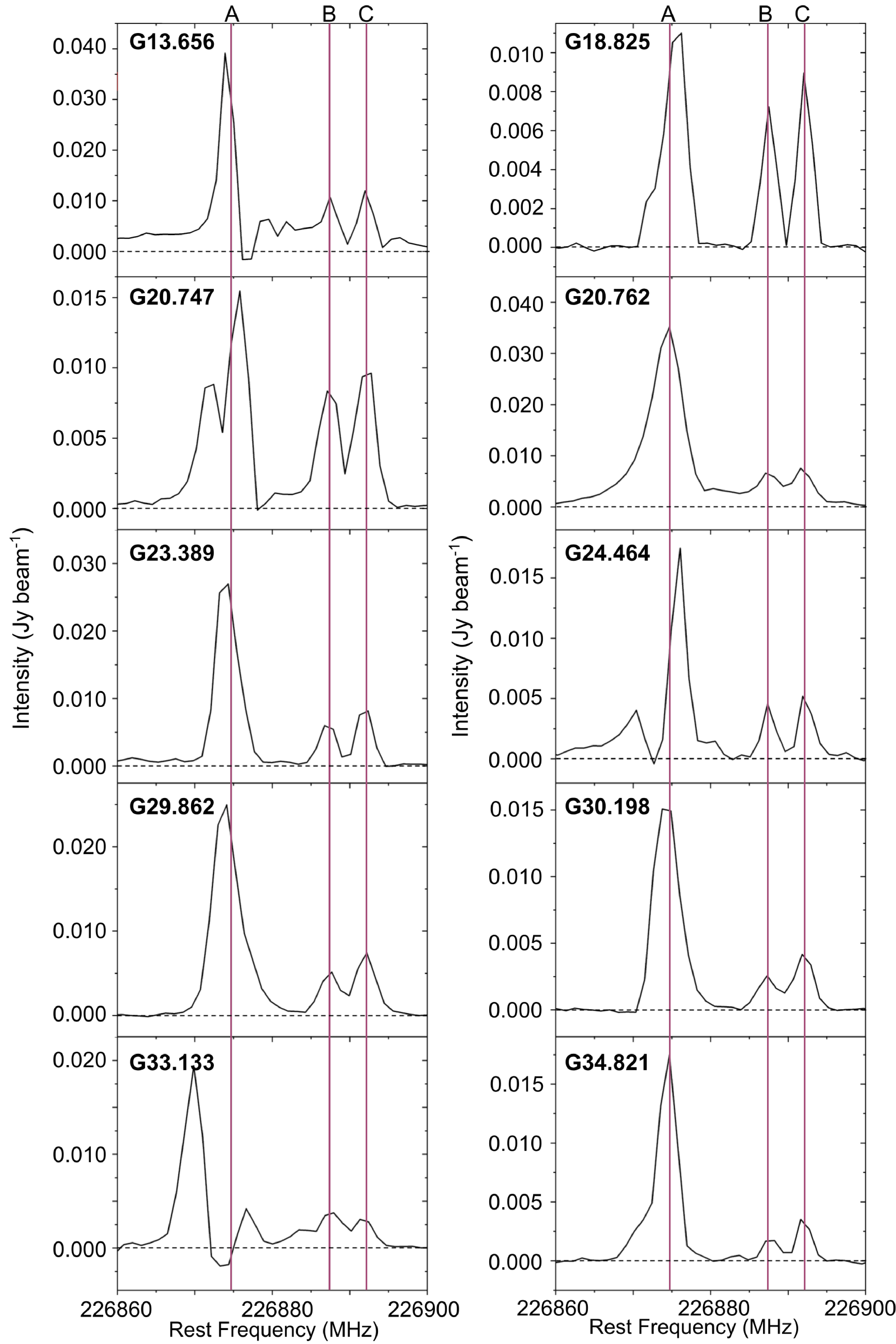}
    \caption{Average spectra obtained from regions of about 5\s~in radius towards each source with  the aim to identify CN lines. 
    The spectra are presented in rest frequency. It is only displayed the frequency range in which transitions of CN appear.
    The vertical lines indicate such transitions: N$=$2--1 J$=$5/2--3/2 F$=$7/2--5/2 (A), N$=$2--1 J$=$5/2--3/2, F$=$3/2--3/2 and F$=$1/2--1/2, (B) and (C), respectively.}
\label{cnLinea}
\end{figure}

To identify CN lines, from the data cubes we extracted spectra of the averaged emission obtained from regions of about 5\s~in radius centered in each source. They are presented in Fig.\,\ref{cnLinea}.
The frequency was converted to rest frequency using the velocities 
listed in Table\,\ref{tabsources}, and it is displayed the frequency range 226860--226900 MHz, in which according to the NIST database, only appears lines of CN, except for one weak line of CH$_{3}$OCHO at 226862.239 MHz. The strongest emission in all cases correspond to the CN N$=$2--1 J$=$5/2--3/2 F$=$7/2--5/2 transition, which as mentioned above, it may be blended with the weaker F$=$5/2--3/2 and F$=$3/2--1/2 lines, but the spectral resolution of the data do not allow us to resolve them properly. Additionally, the CN lines N$=$2--1 J$=$5/2--3/2, F$=$3/2--3/2 and F$=$1/2--1/2 at 226887.399 and 226892.151 MHz, respectively, appear clearly in all spectra. In some cases (see spectra of sources G13.656, G20.747, G24.464, and G33.133), the main component presents probably absorption or self-absorption features. This will be specially discussed below. 

\begin{figure}[h!]
    \centering
    \includegraphics[width=4.459cm]{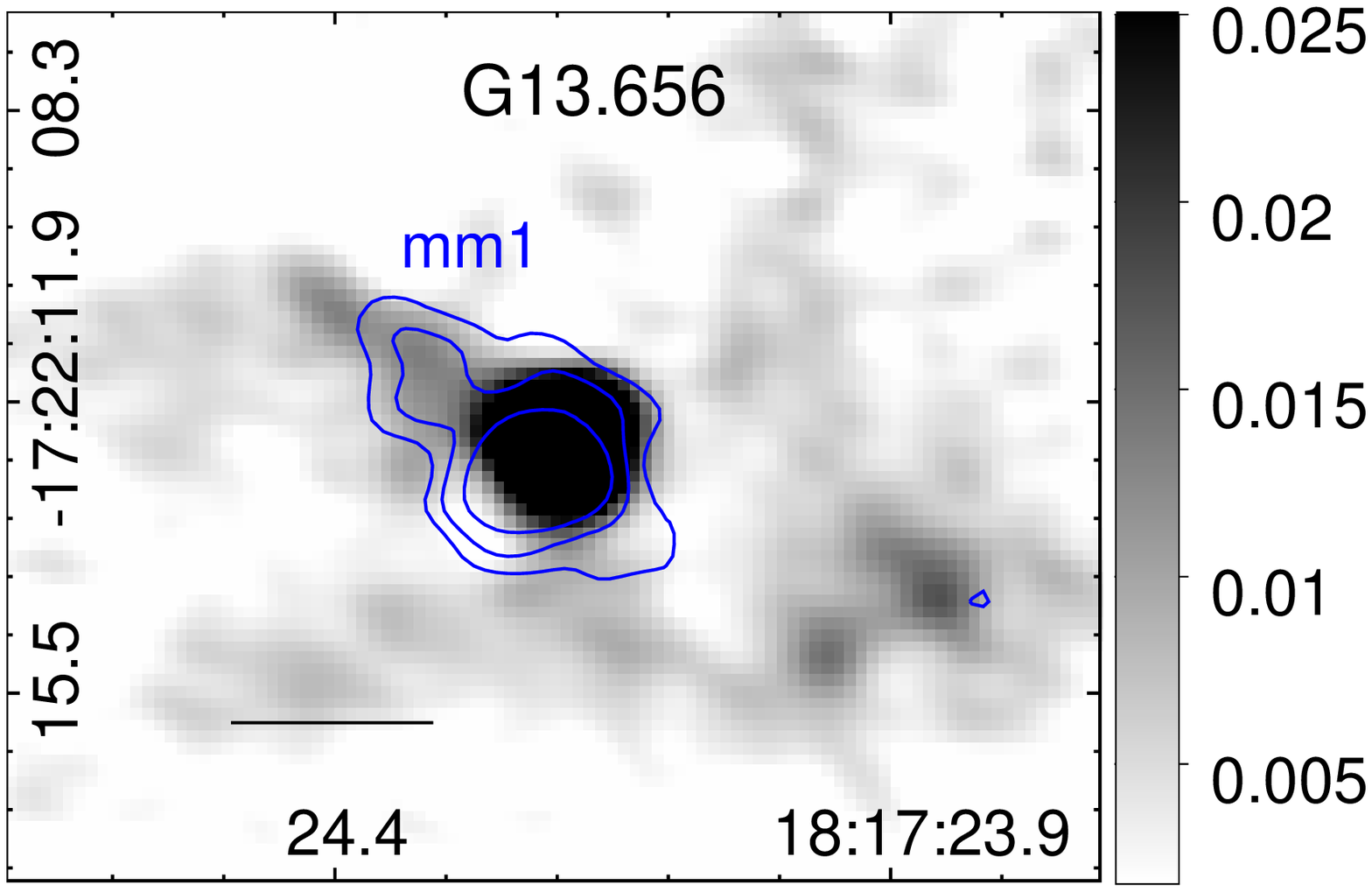}
    \includegraphics[width=4.45cm]{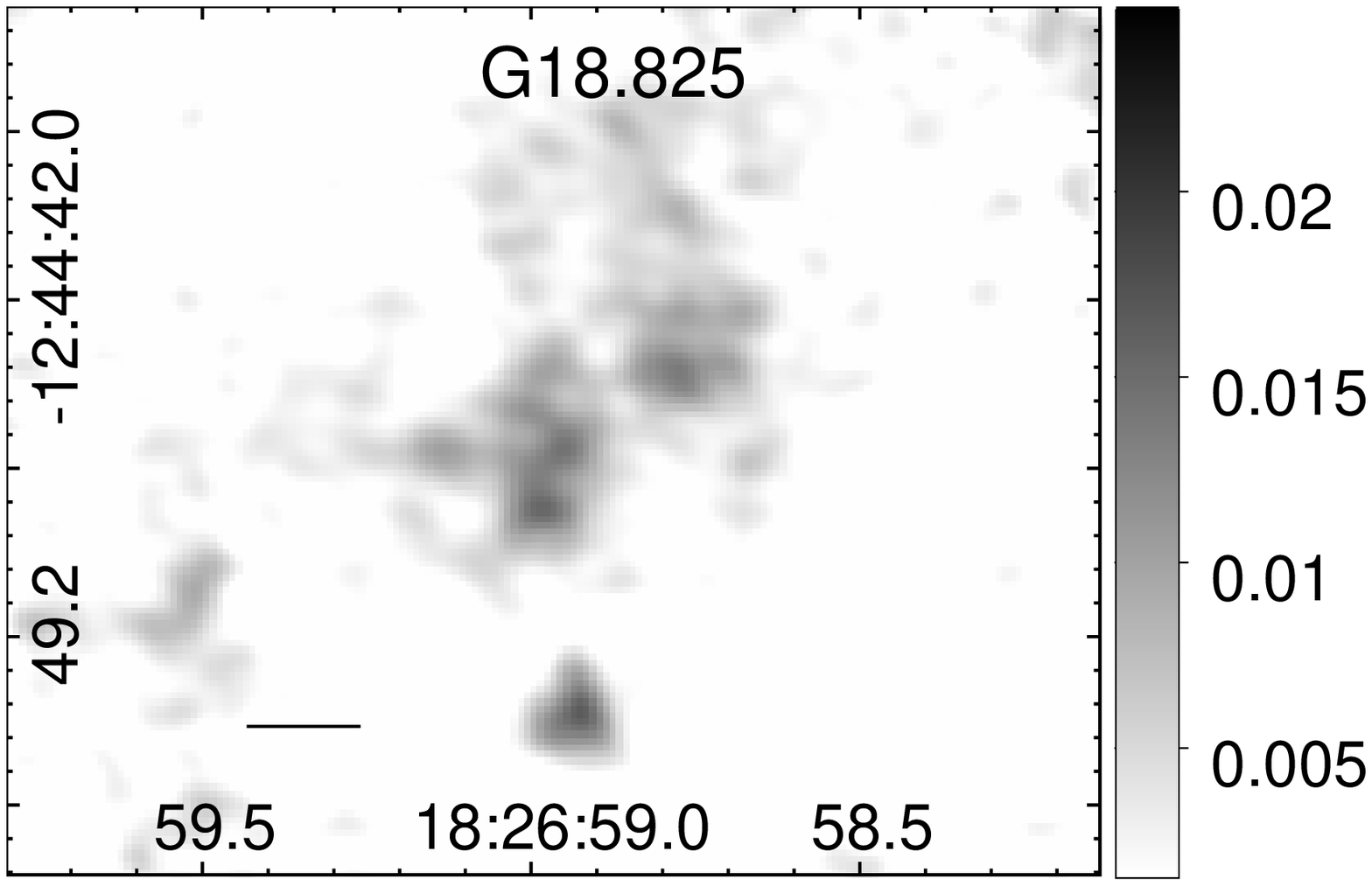}
    \includegraphics[width=4.459cm]{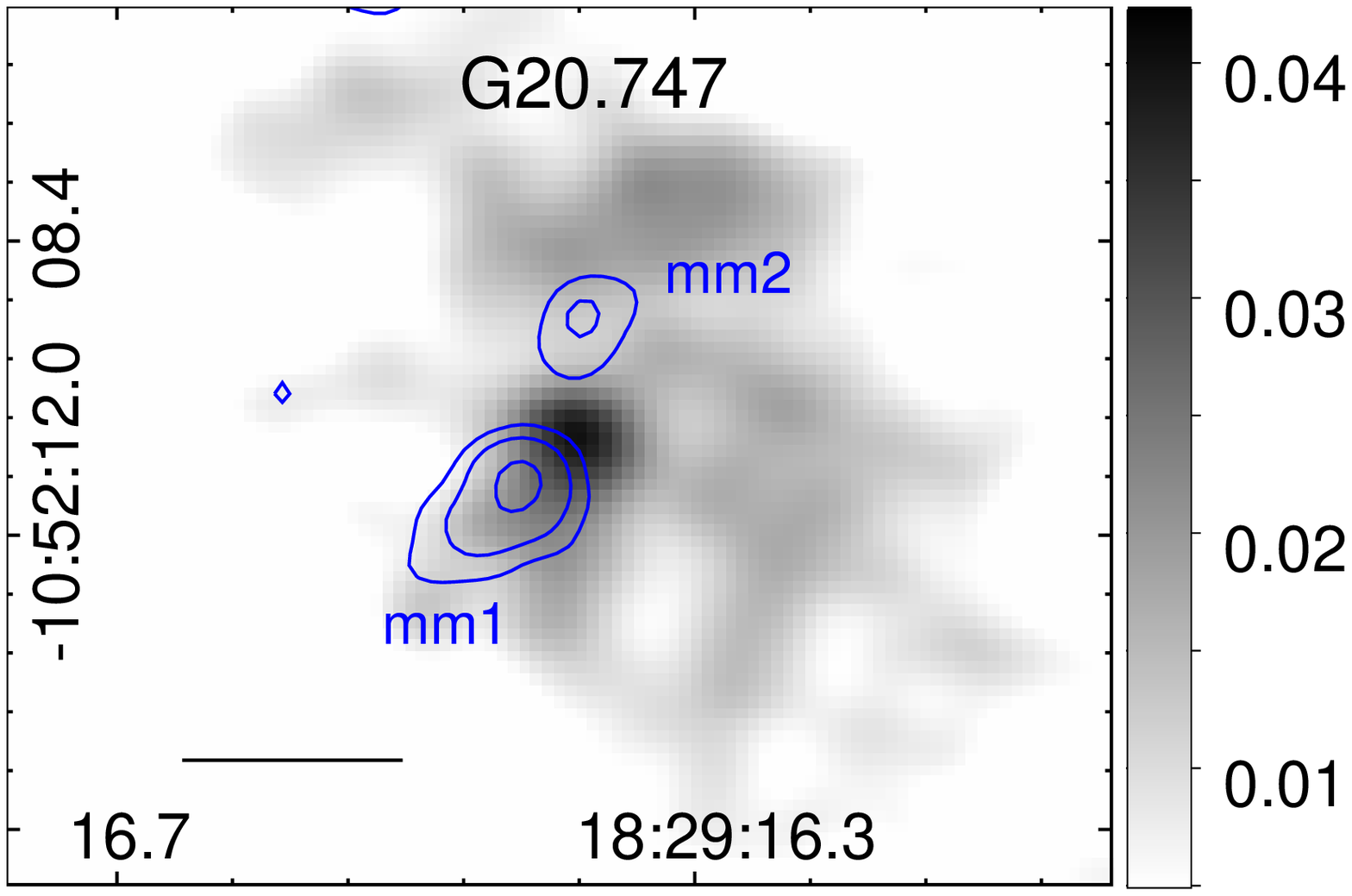}
    \includegraphics[width=4.459cm]{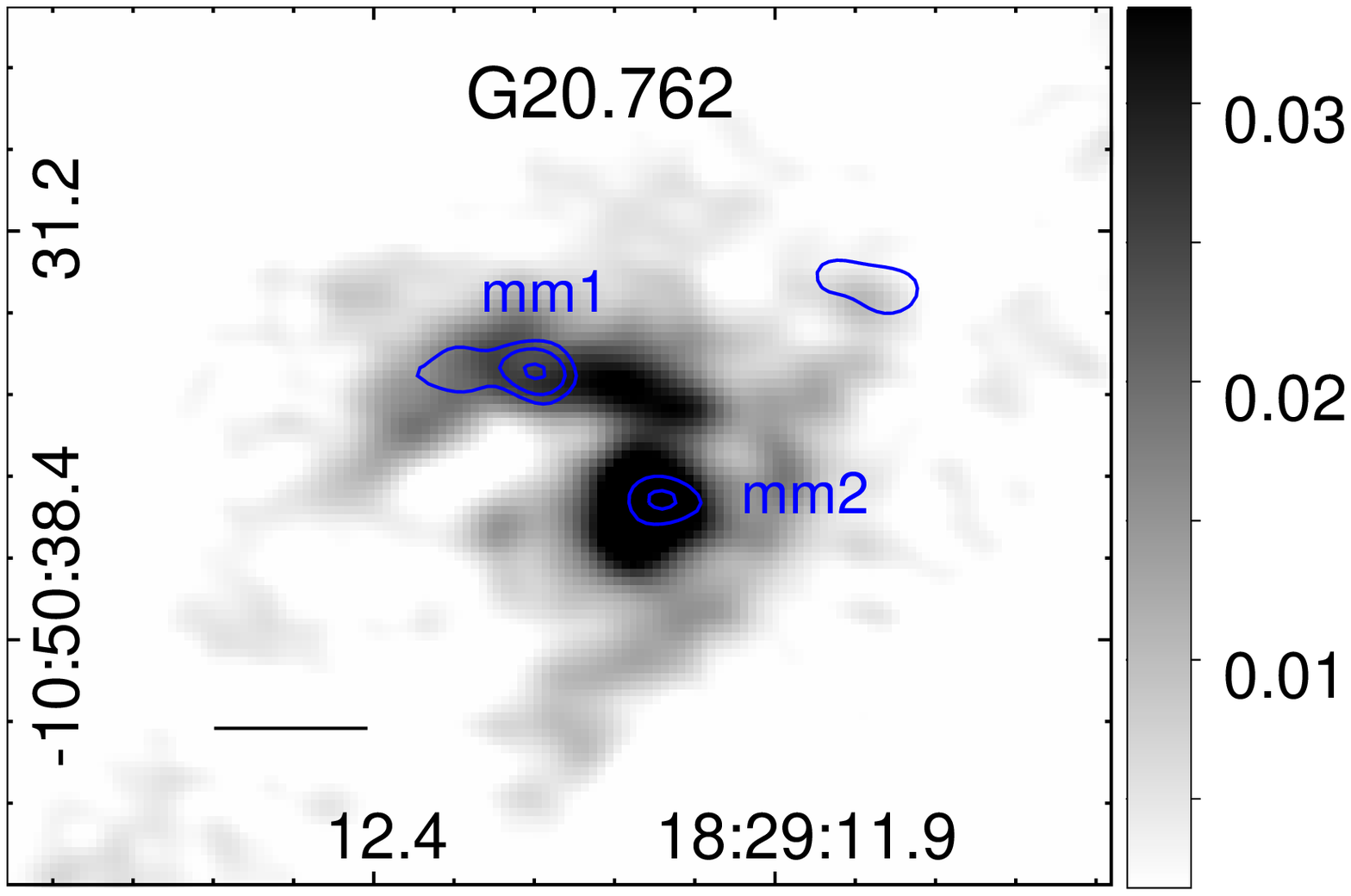}
    \includegraphics[width=4.459cm]{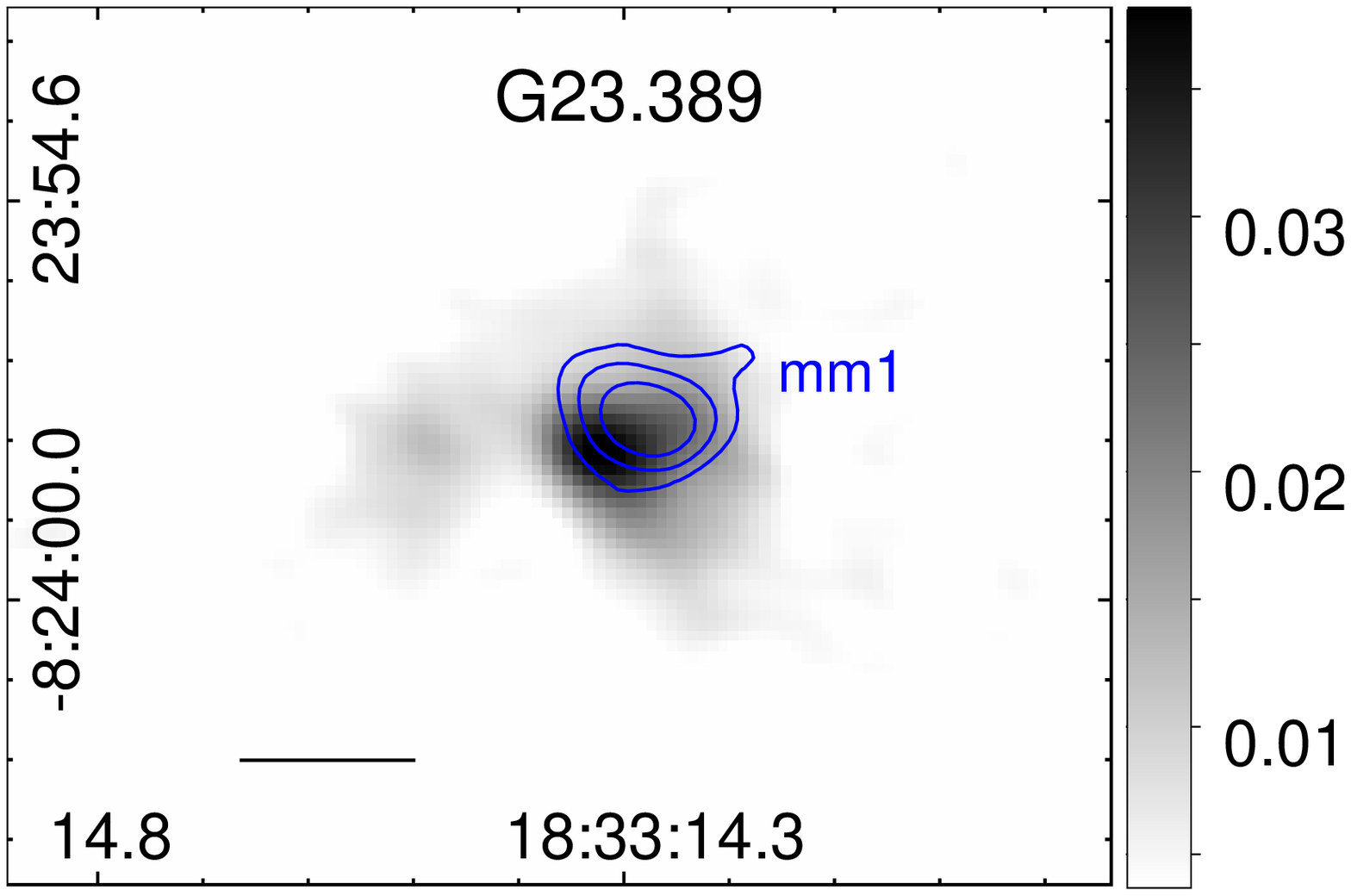}
    \includegraphics[width=4.459cm]{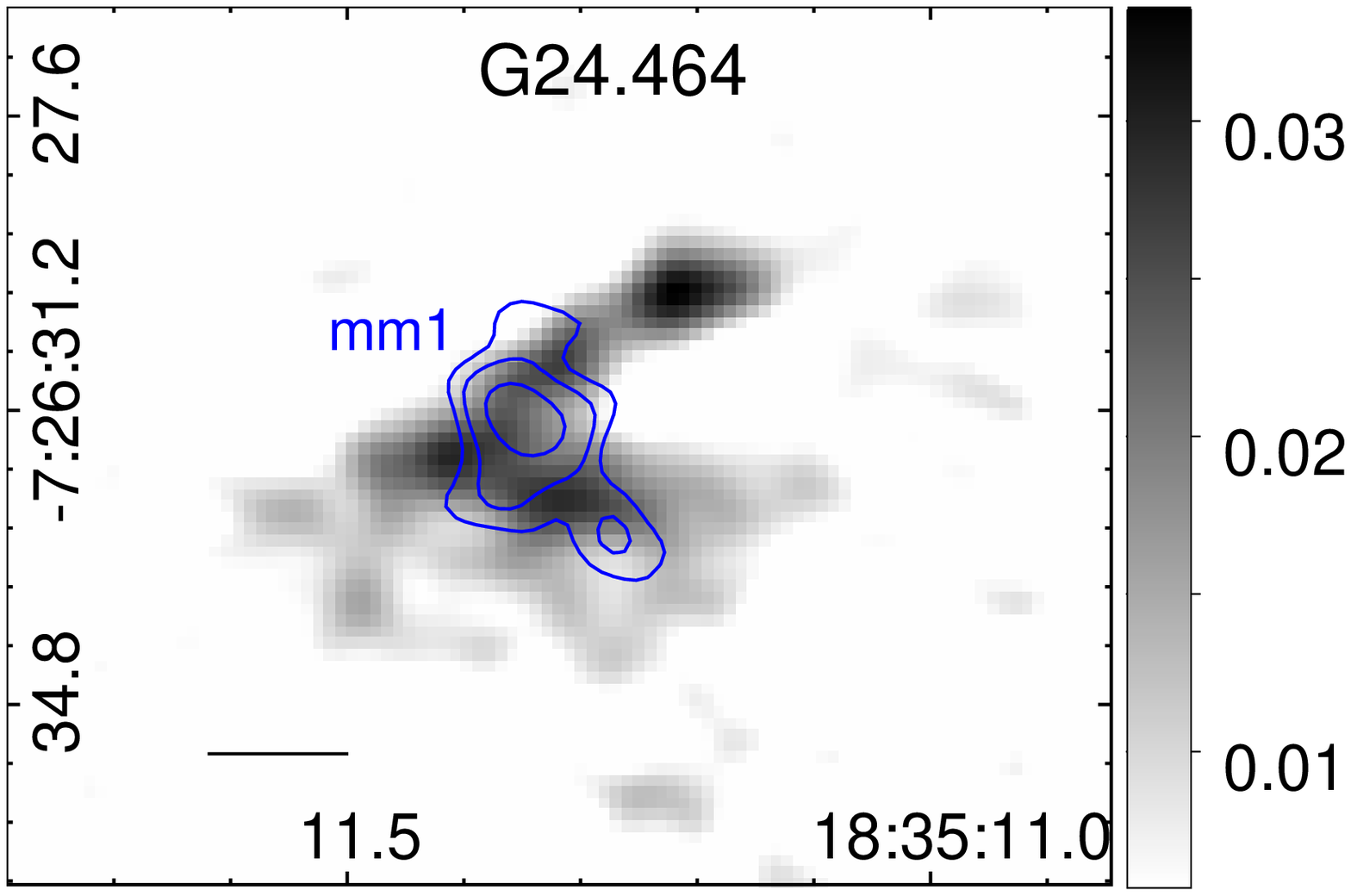}
    \includegraphics[width=4.459cm]{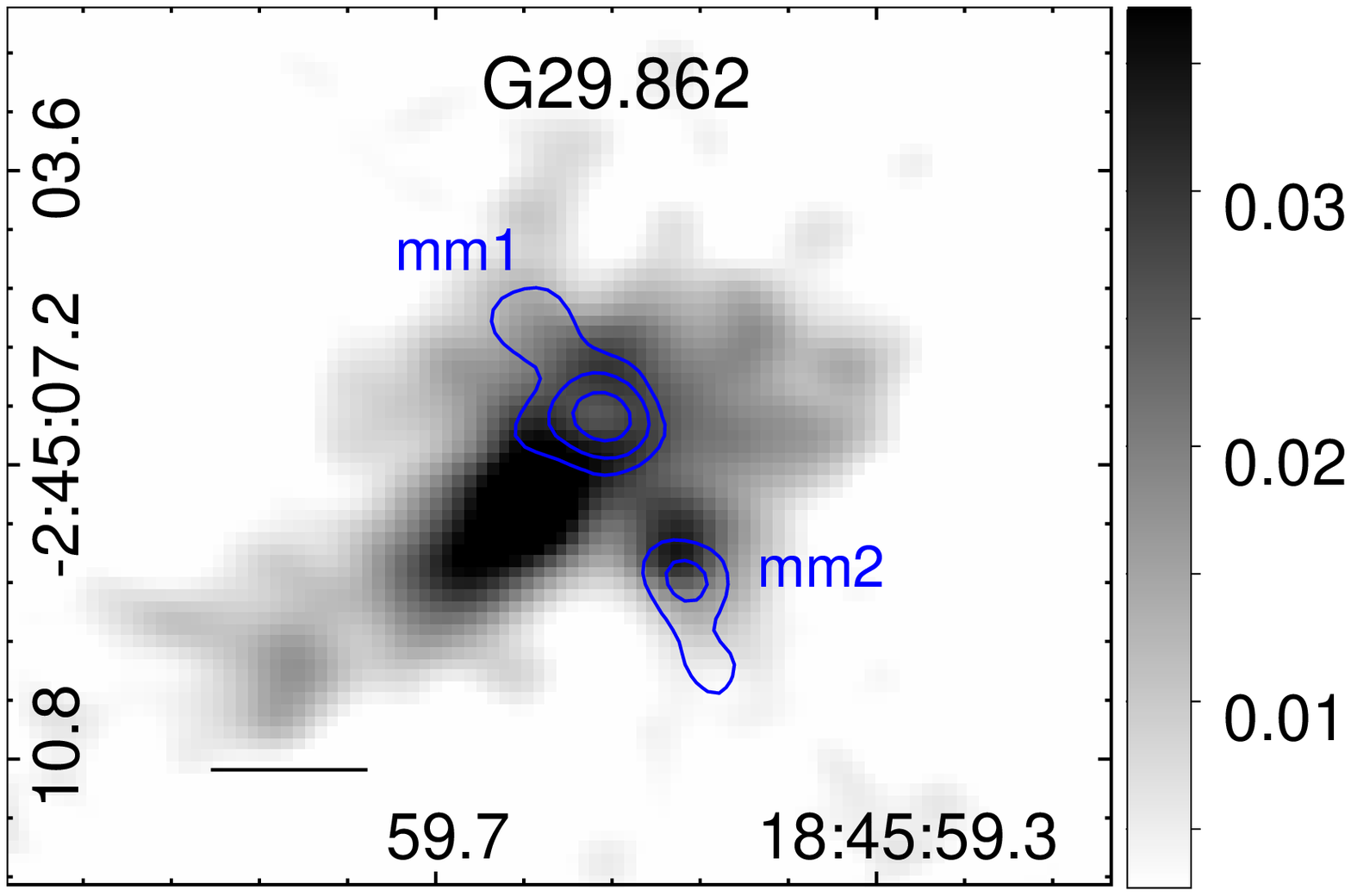}
    \includegraphics[width=4.459cm]{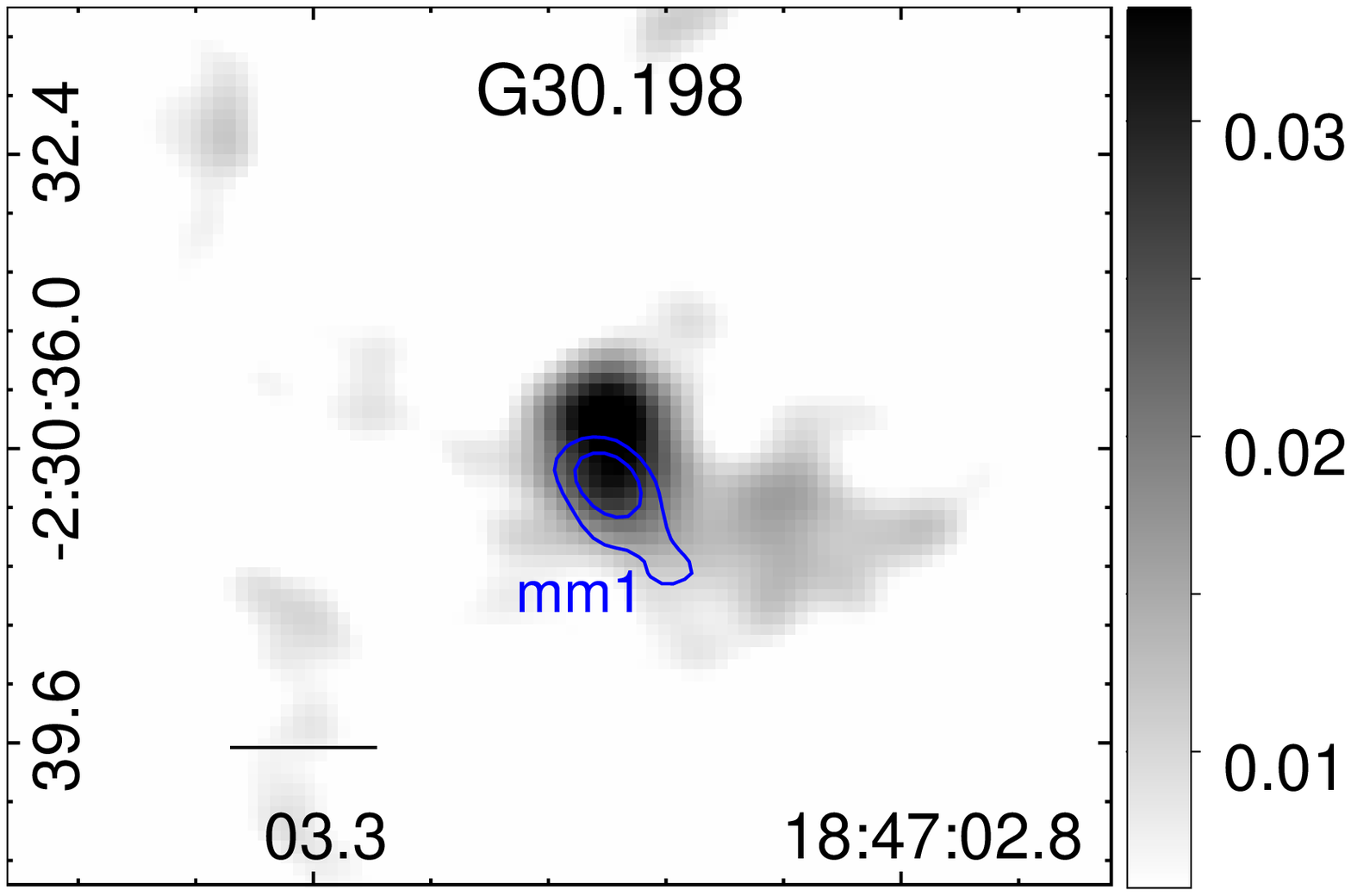}
    \includegraphics[width=4.459cm]{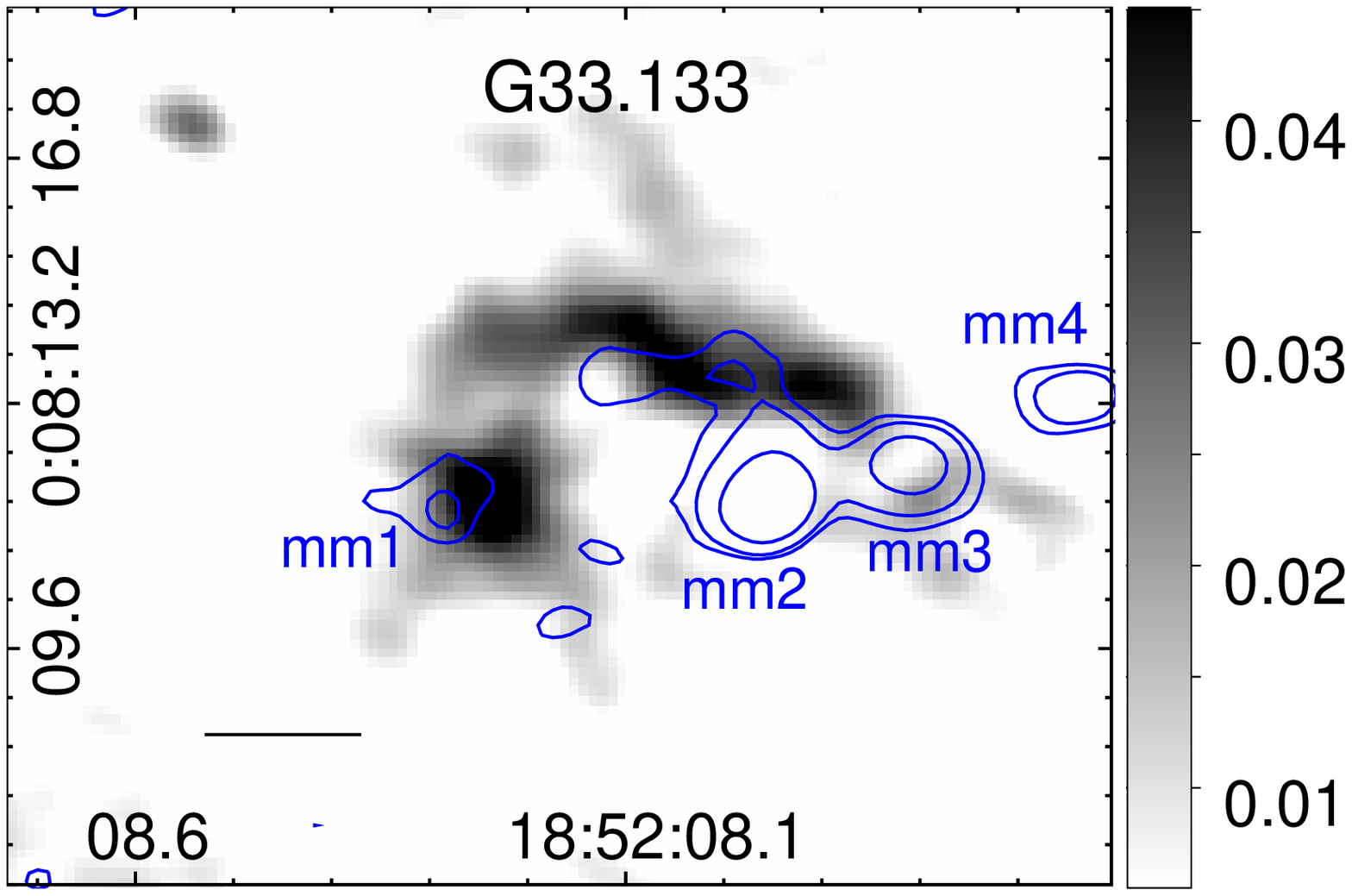}
    \includegraphics[width=4.459cm]{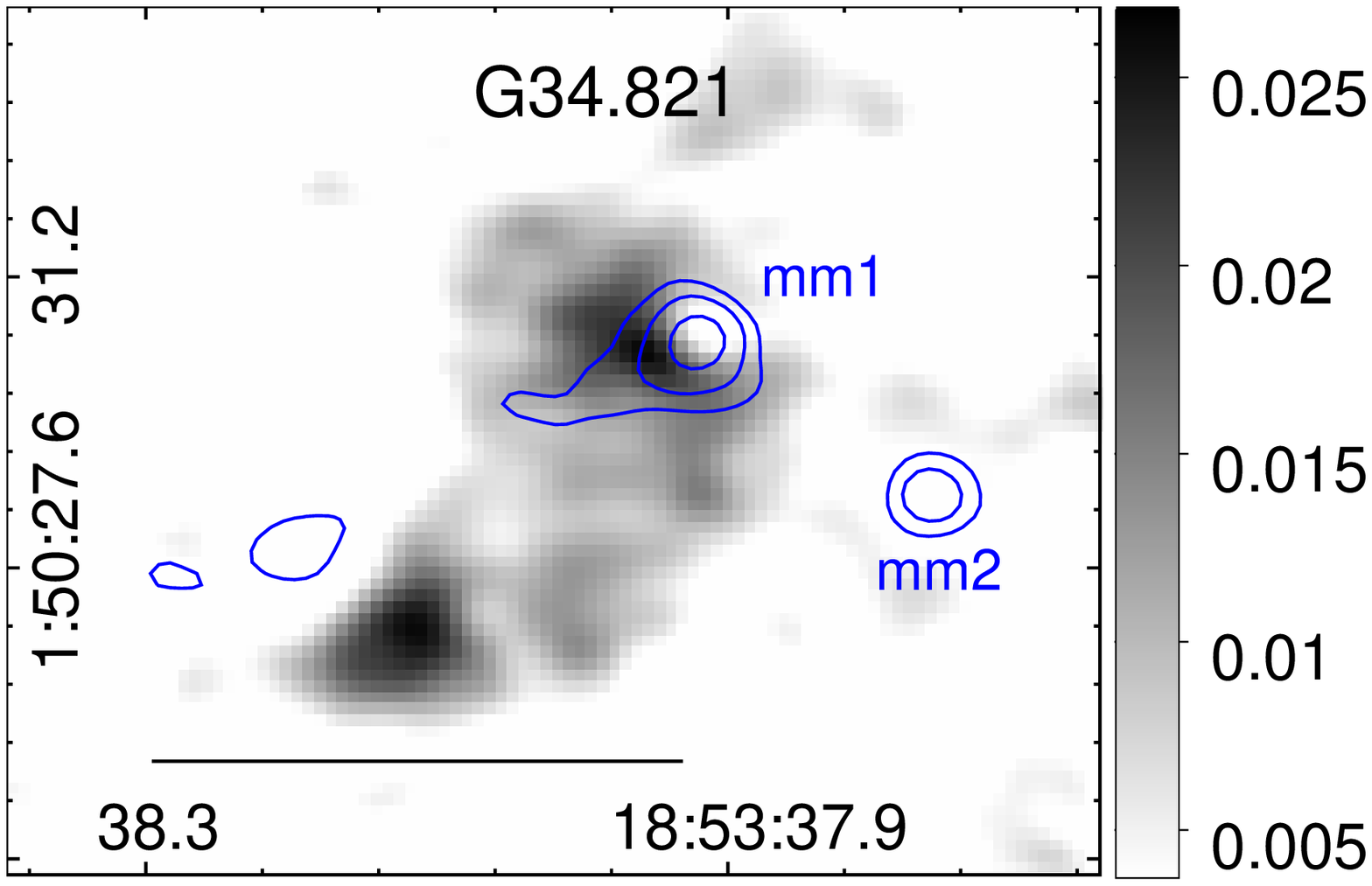}
    \caption{Maps of the CN emission averaged along the frequency interval of the main component. The units of the colorbar is Jy beam$^{-1}$. The blue contours show the continuum emission at 1.3 mm with levels of 0.005, 0.010, and 0.020 Jy beam$^{-1}$, and the names of the millimeter source used in Table\,\ref{contparams} are indicated in each case. The 1$\sigma$ rms noise level is about 0.0015 Jy beam$^{-1}$ for both the continuum and the averaged line emission. The horizontal lines at the bottom left corner in each panel represent a size of 0.05 pc for each case.}
\label{cnAve}
\end{figure}

To appreciate the spatial distribution of the CN emission, we present in Fig.\,\ref{cnAve} maps of the 
CN emission averaged along the frequency interval of the main component. Additionally, the continuum emission at
1.3 mm is displayed in contours. Source G18.825 does not present continuum emission above the noise level. 
Given that we are presenting here the spectrum of the millimeter emission of source G29.862, we point out that in a previous work, in which we performed a detailed multiwavelength study of this source (see \citealt{areal20}), the CN line was misinterpreted as CH$_3$OCHO (erratum in preparation).

Table\,\ref{contparams}~presents the parameters of the cores observed at 1.3 mm obtained from a two dimensional Gaussian fitting. The peak intensity and the integrated flux density corrected for primary beam response are presented in Cols.\,2 and\,3, the source size in arcsec deconvolved from beam is presented in Col.\,4, and in Col.\,5, a rough estimate of the core size in astronomical units considering the average between the size of the semi-axes presented in Col.\,4 and the distances listed in Table\,\ref{tabsources}. The source of the errors in the size is mainly the uncertainties in the distances, and it can be about 20$\%$.

\begin{table}
\caption{Parameters of the 1.3 mm continuum cores indicated in Fig.\,\ref{cnAve}.}
\label{contparams}
\tiny
\centering
\begin{tabular}{lcccc}
\hline\hline
Core    & $I_{peak}$        & $S$          & $\theta_{s}$          &  Size    \\
          & (mJy beam$^{-1}$)        & (mJy)        &   (\s)                &    au     \\
\hline                 
G13.656-mm1   &  114.0$\pm$5.7    & 224$\pm$16    & 0.82$\times$0.64  & 3000 \\
G20.747-mm1  &  22.2$\pm$1.4     & 69.6$\pm$5.8  & 1.48$\times$0.73  & 4100 \\
G20.747-mm2  &  10.4$\pm$1.0     & 25.5$\pm$3.4  & 1.27$\times$0.48  & 3400 \\
G20.762-mm1  &  17.1$\pm$2.2     & 43.9$\pm$7.8  & 1.87$\times$0.28  & 4200 \\
G20.762-mm2  &  11.5$\pm$1.0     & 18.7$\pm$2.6  & 0.87$\times$0.36  & 2400 \\
G23.389-mm1  &  44.4$\pm$2.1     & 91.5$\pm$6.0  & 0.91$\times$0.68  & 3500 \\ 
G24.464-mm1  &  24.9$\pm$2.4     & 112$\pm$13    & 1.63$\times$1.04  & 8300 \\
G29.862-mm1  &  26.2$\pm$1.9     & 51.6$\pm$5.3  & 0.77$\times$0.66  & 4000 \\   
G29.862-mm2  &  10.0$\pm$1.1     & 30.4$\pm$4.4  & 0.80$\times$0.39  & 3300  \\
G30.198-mm1  &  12.6$\pm$1.5     & 40.2$\pm$6.1  & 1.60$\times$0.62  & 6500 \\
G33.133-mm1  &  11.4$\pm$1.6     & 30.3$\pm$5.8  & 1.15$\times$0.72  & 4200 \\
G33.133-mm2  &  102.4$\pm$4.9    & 207$\pm$14    & 0.85$\times$0.56  & 3100 \\
G33.133-mm3  &  67.8$\pm$2.9     & 113.1$\pm$7.1 & 0.72$\times$0.47  & 2600 \\
G33.133-mm4  &  22.2$\pm$1.2     & 35.6$\pm$3.0  & 0.85$\times$0.28  & 2500 \\
G34.821-mm1  &  24.7$\pm$2.4     & 64.0$\pm$8.4  & 1.03$\times$0.81  & 1500 \\
G34.821-mm2  &  16.1$\pm$0.4     & 21.7$\pm$0.8  & 0.49$\times$0.32  & 650 \\
\hline
\end{tabular}
\end{table}

\begin{figure}[h!]
    \centering
    \includegraphics[width=4.45cm]{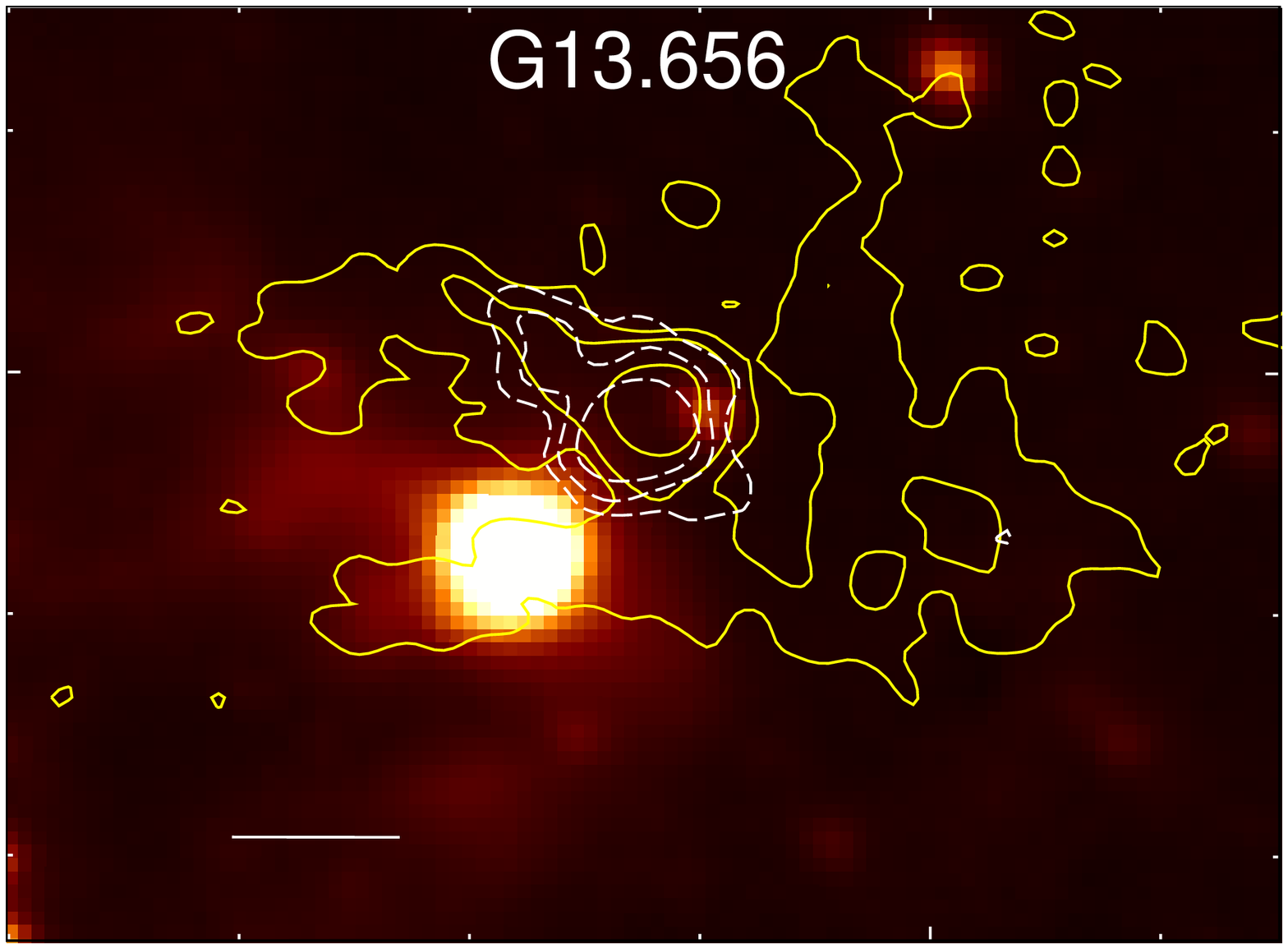}
    \includegraphics[width=4.45cm]{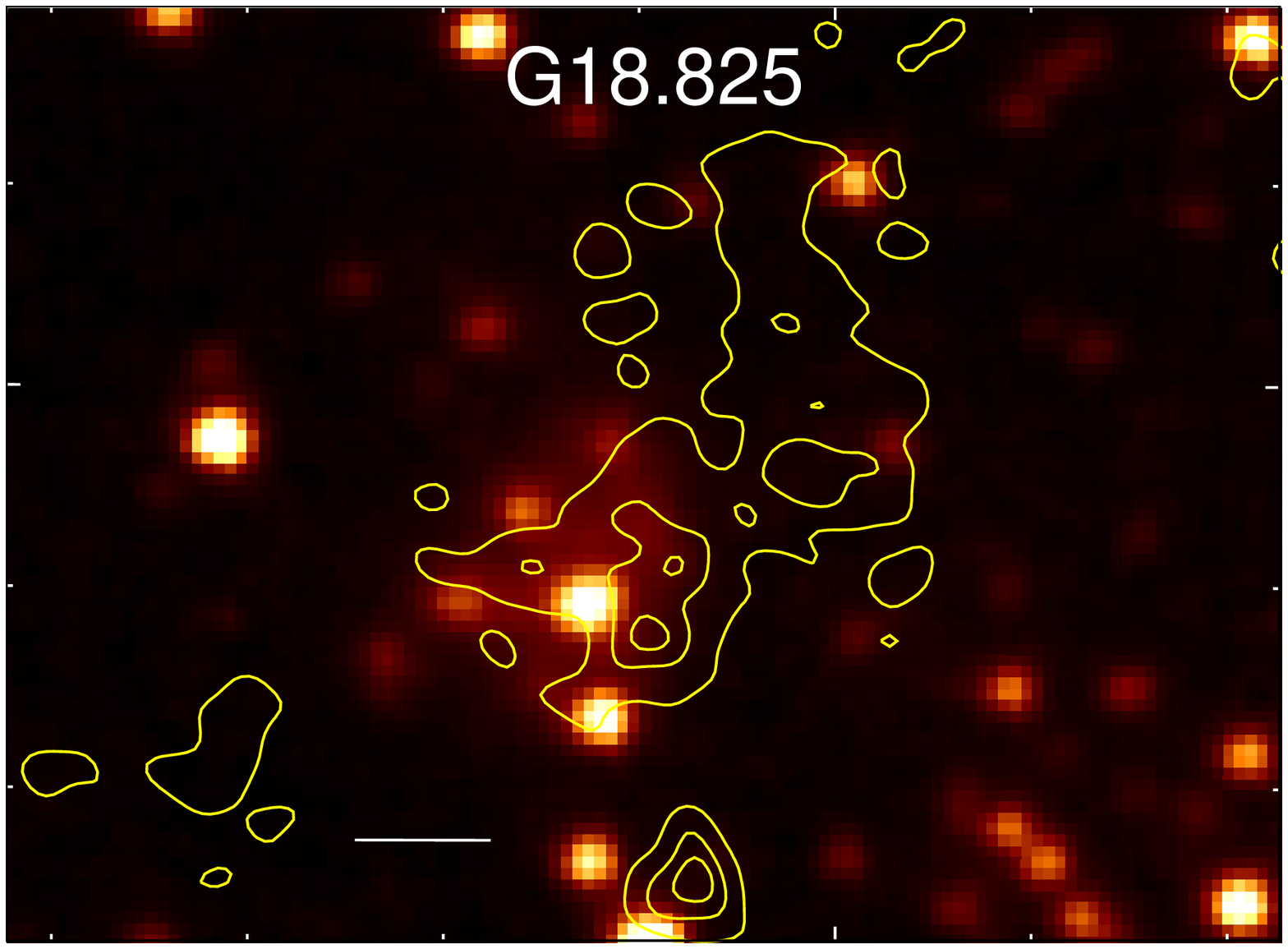}
    \includegraphics[width=4.45cm]{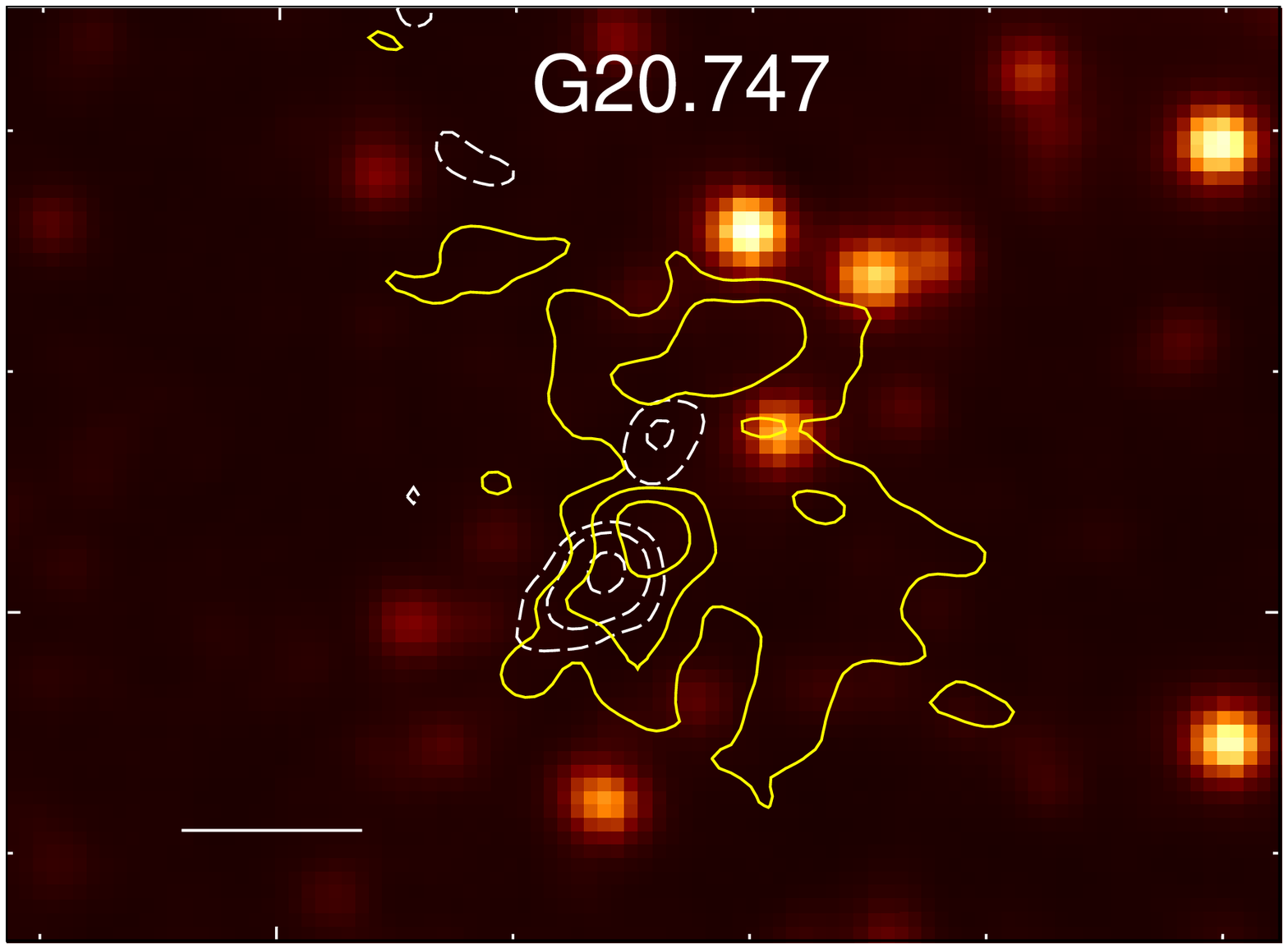}
    \includegraphics[width=4.45cm]{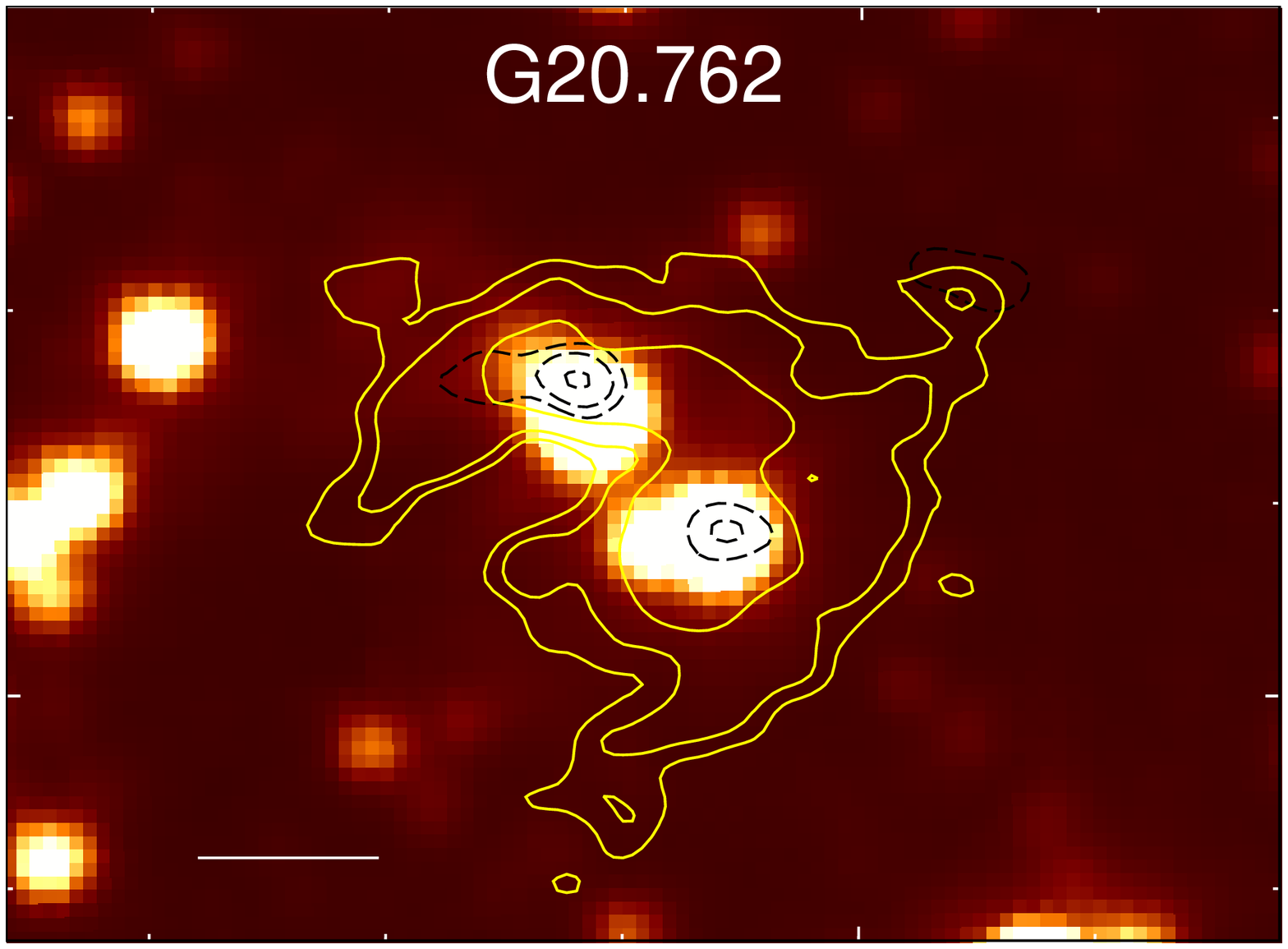}
    \includegraphics[width=4.45cm]{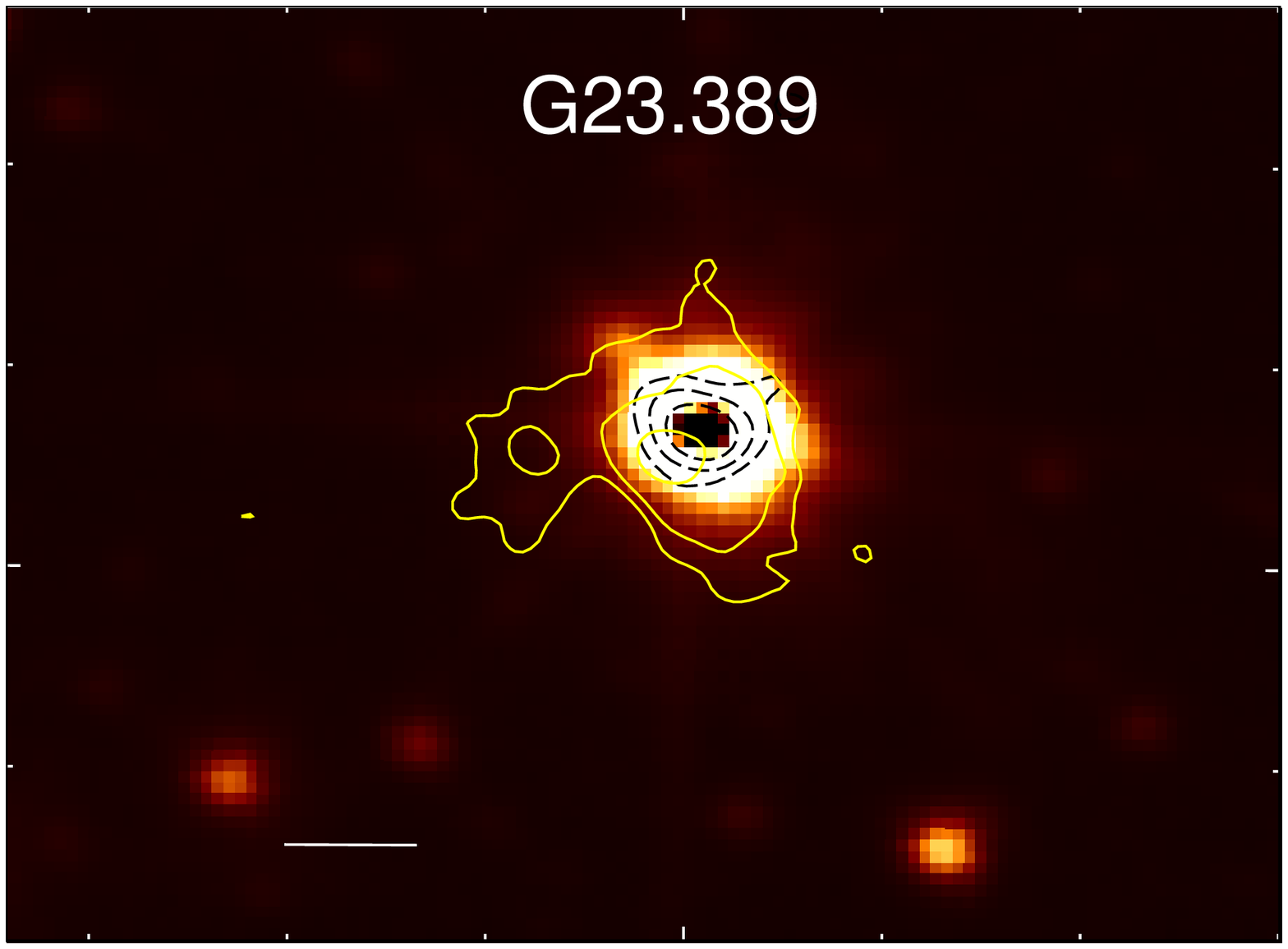}
    \includegraphics[width=4.45cm]{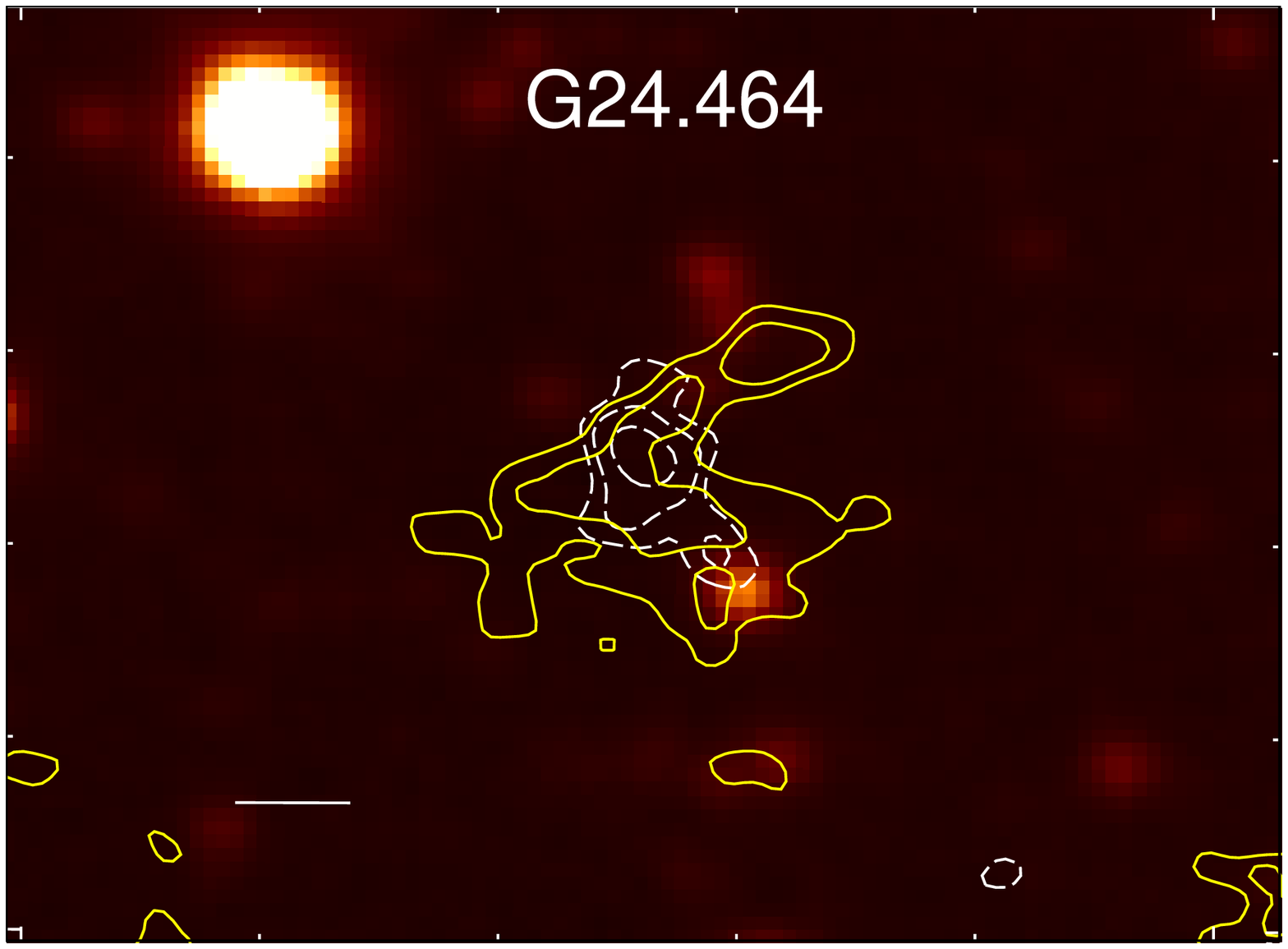}
    \includegraphics[width=4.45cm]{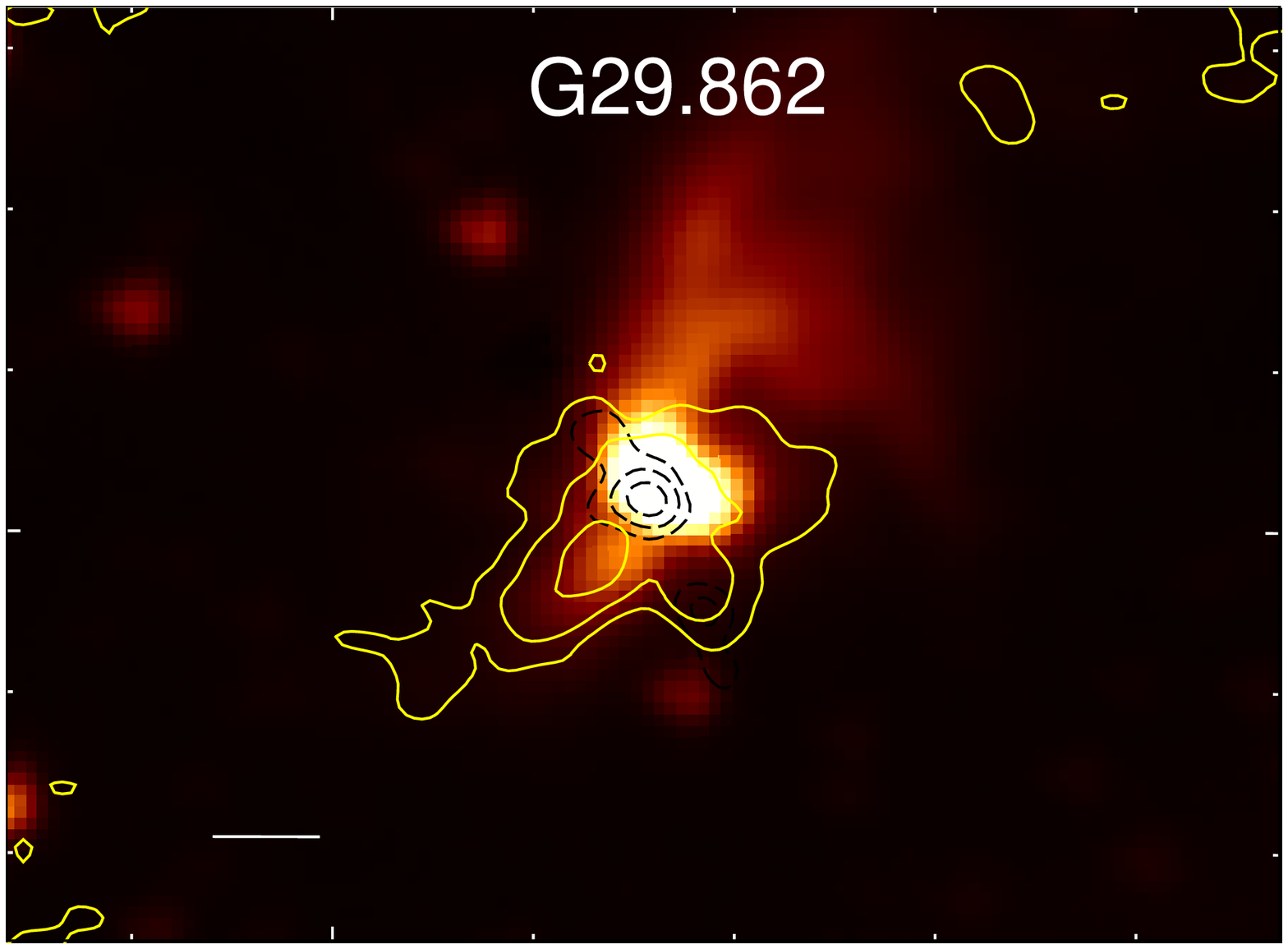}
    \includegraphics[width=4.45cm]{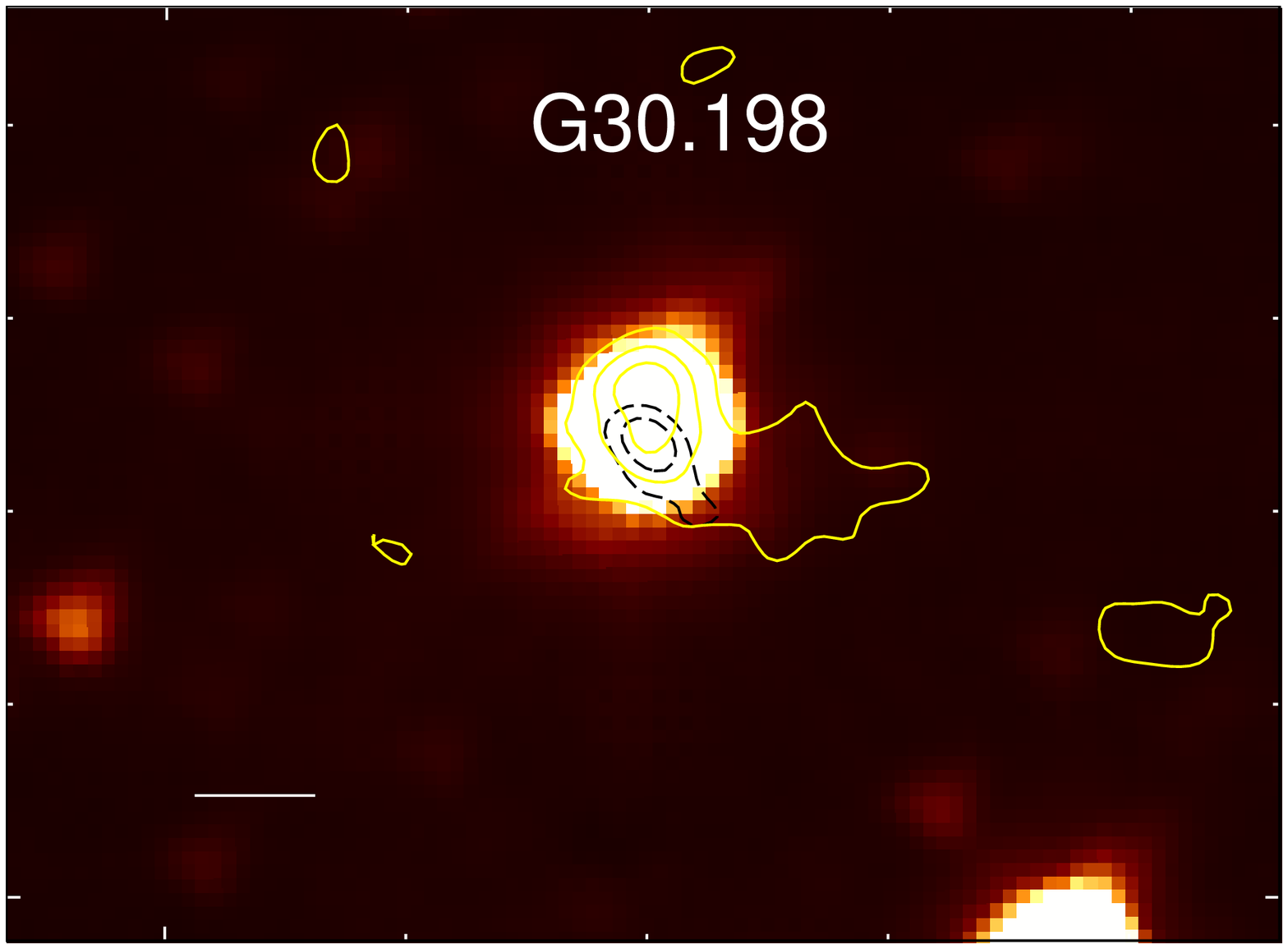}
    \includegraphics[width=4.45cm]{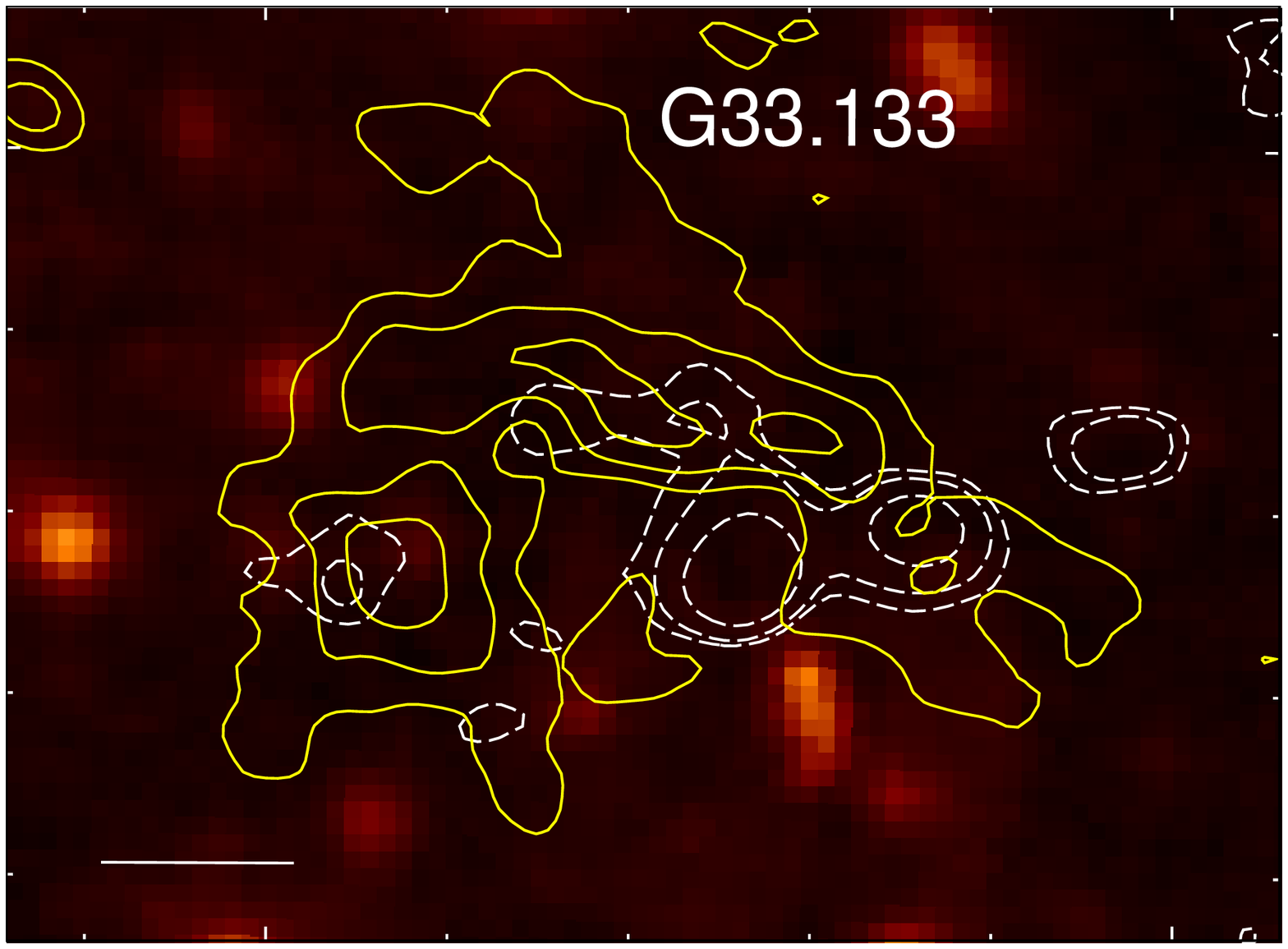}
    \includegraphics[width=4.45cm]{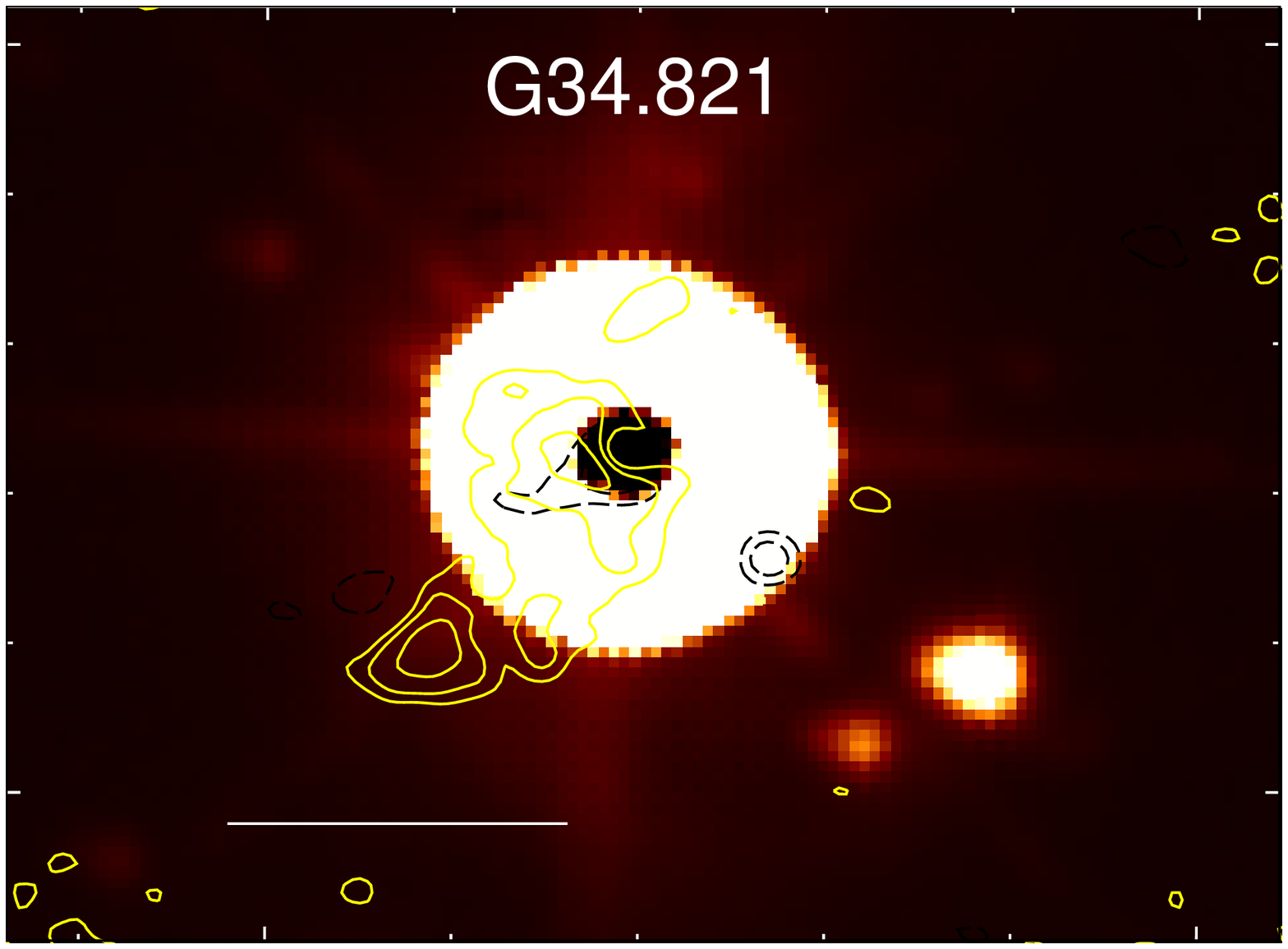} 
    \caption{Near-IR emission at {\it Ks}-band obtained from the UKIDSS database. Yellow contours are the CN 
    averaged emission, and doted contours are the continuum emission at 1.3 mm, both as presented in Fig.\,\ref{cnAve}. In all cases the first contour of the averaged CN emission is between 3 and 4 $\sigma$ 
    rms noise level, and the increase is about 0.01 Jy beam$^{-1}$. The horizontal white lines in each panel represent a size of 0.05 pc for each case.}
\label{cnIR}
\end{figure}

\begin{figure}[h!]
    \centering
    \includegraphics[width=4.3cm]{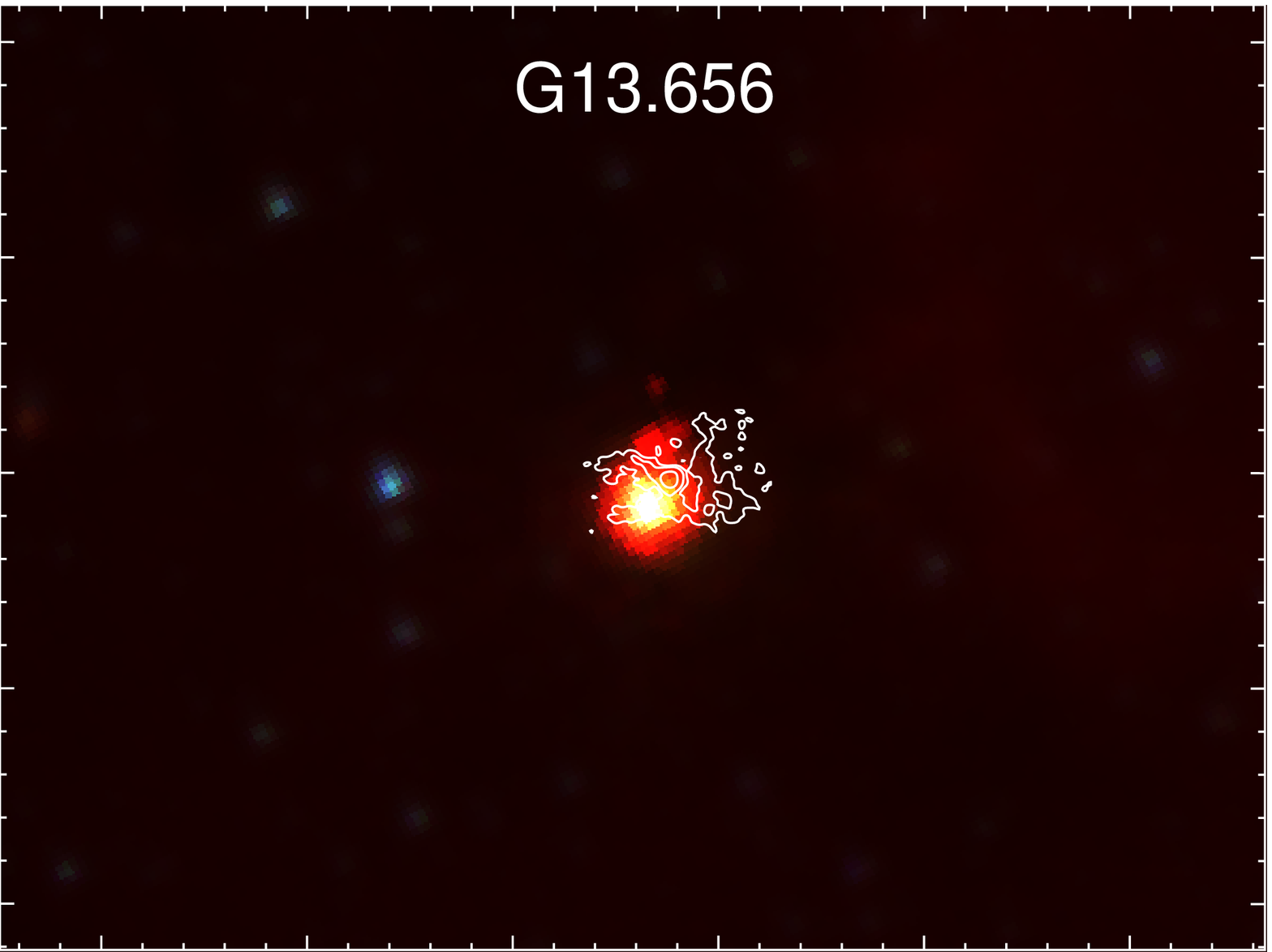}
    \includegraphics[width=4.3cm]{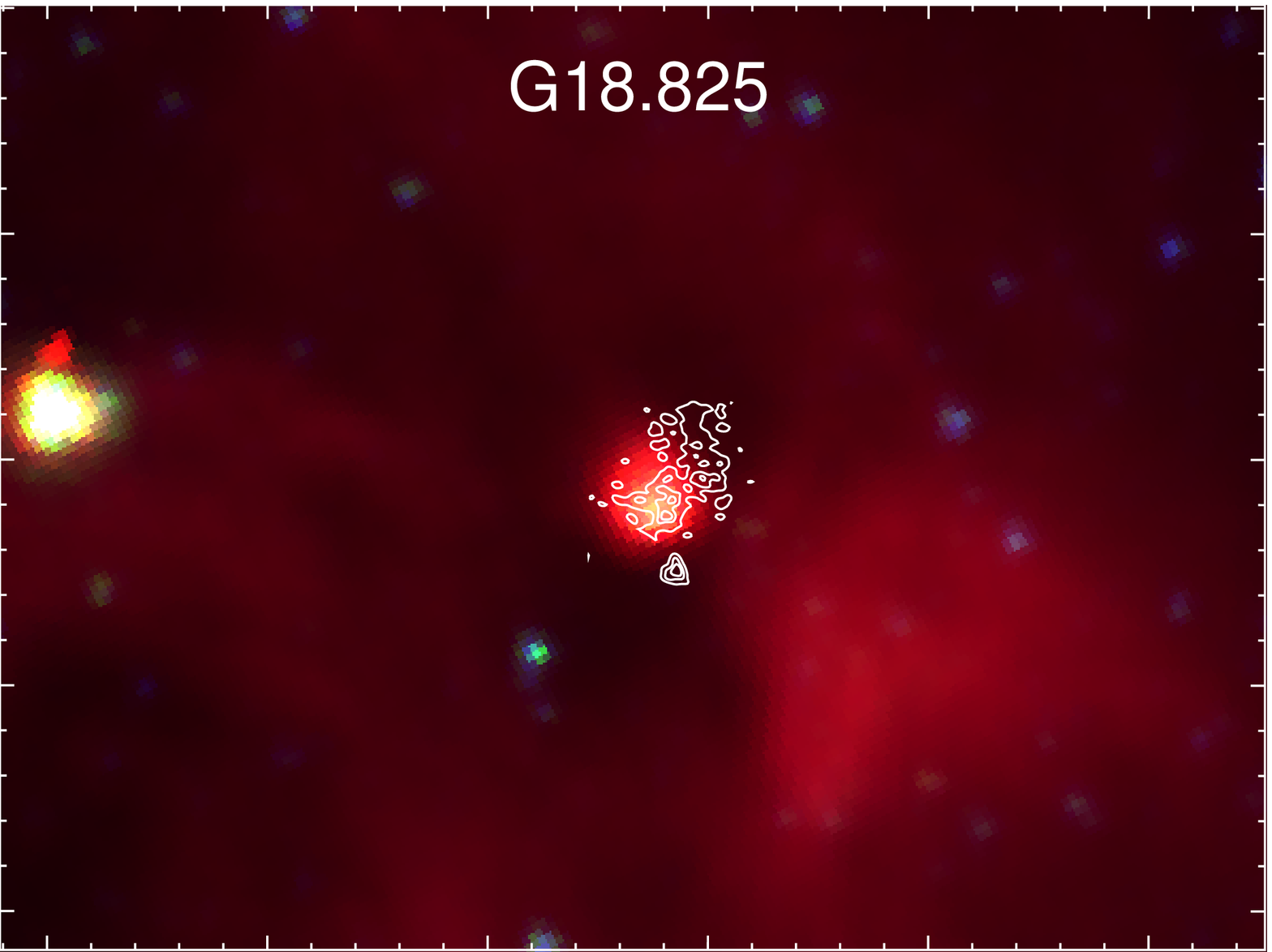}
    \includegraphics[width=4.3cm]{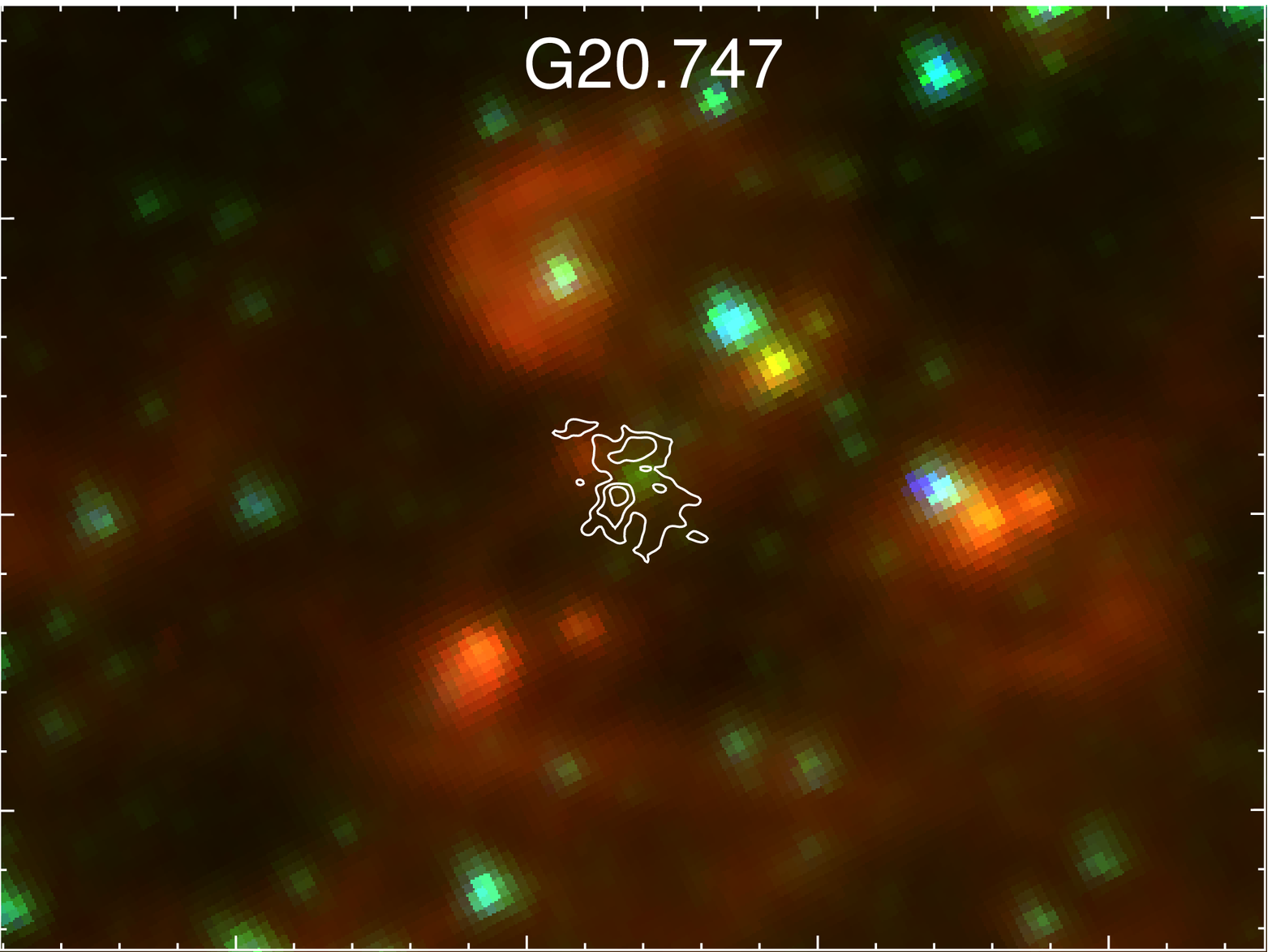}
    \includegraphics[width=4.3cm]{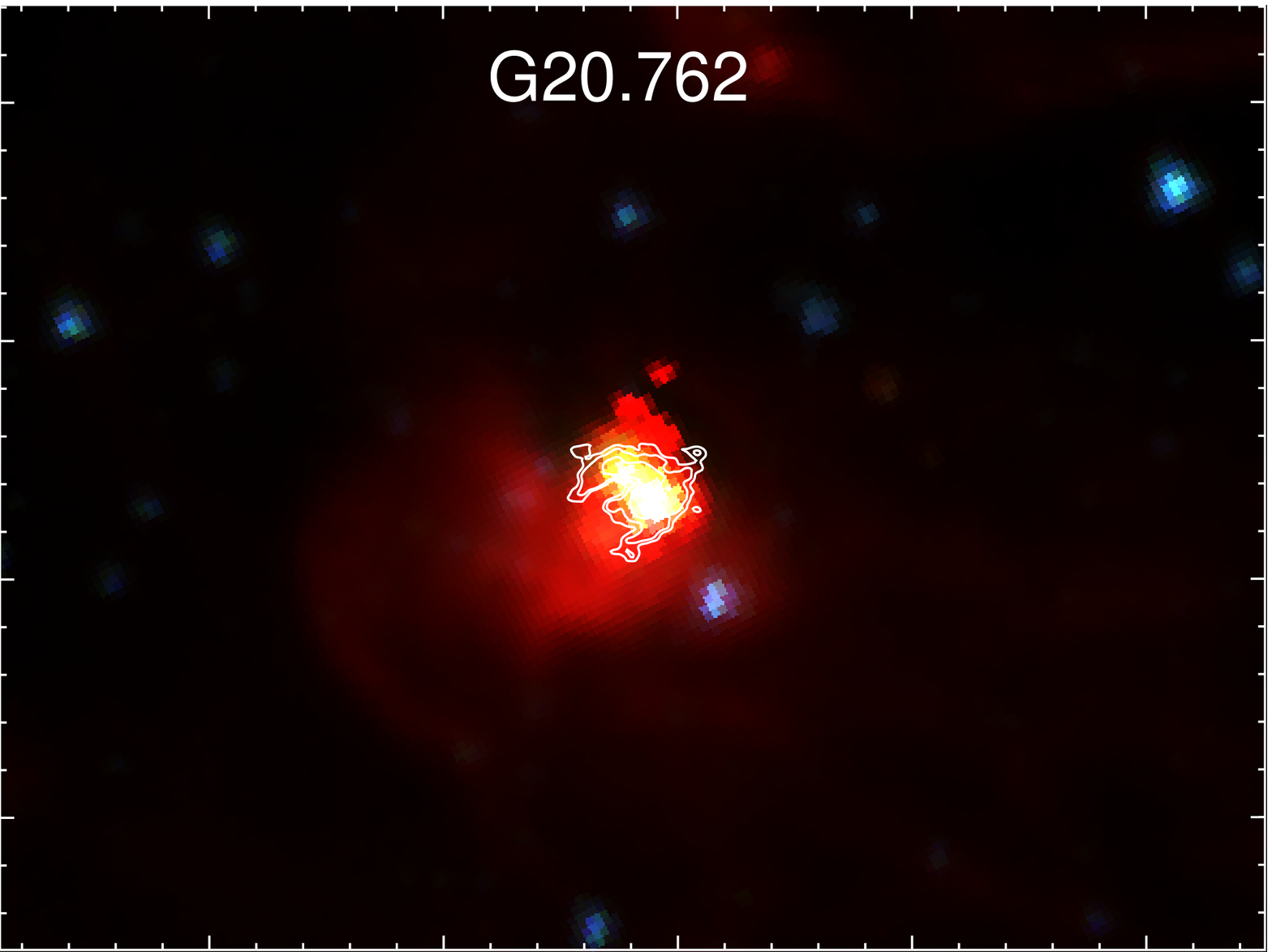}
    \includegraphics[width=4.3cm]{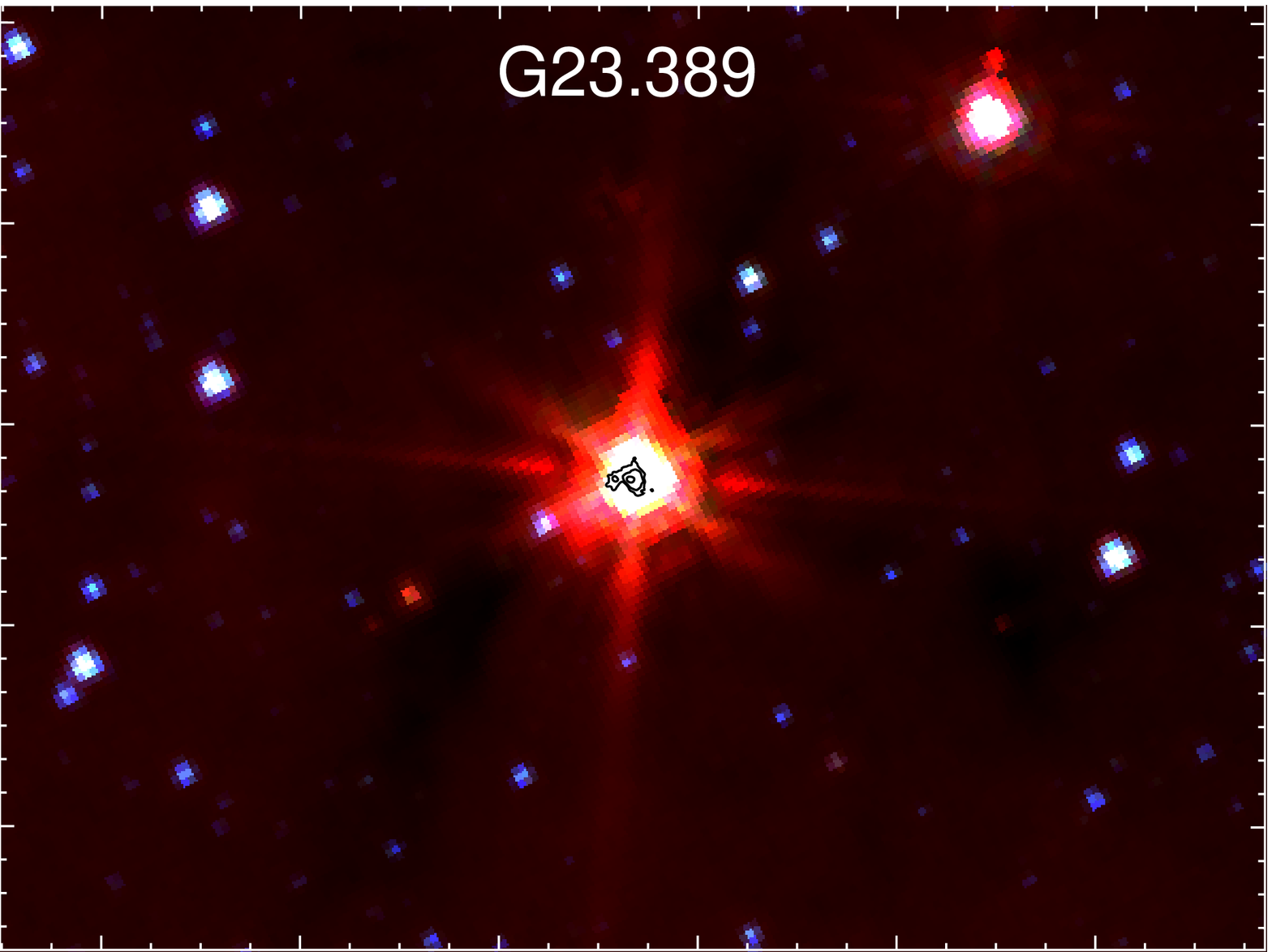}
    \includegraphics[width=4.3cm]{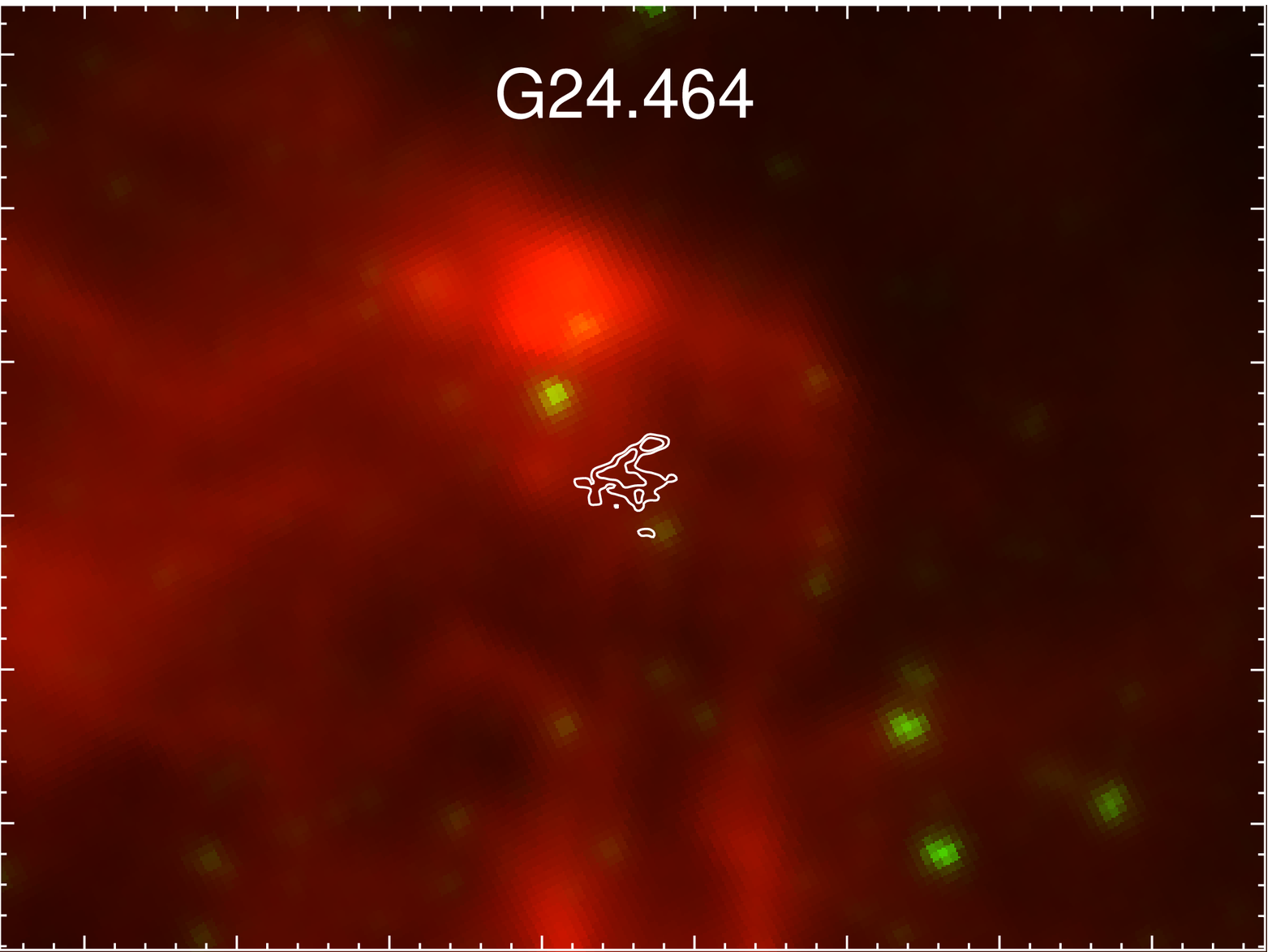}
    \includegraphics[width=4.3cm]{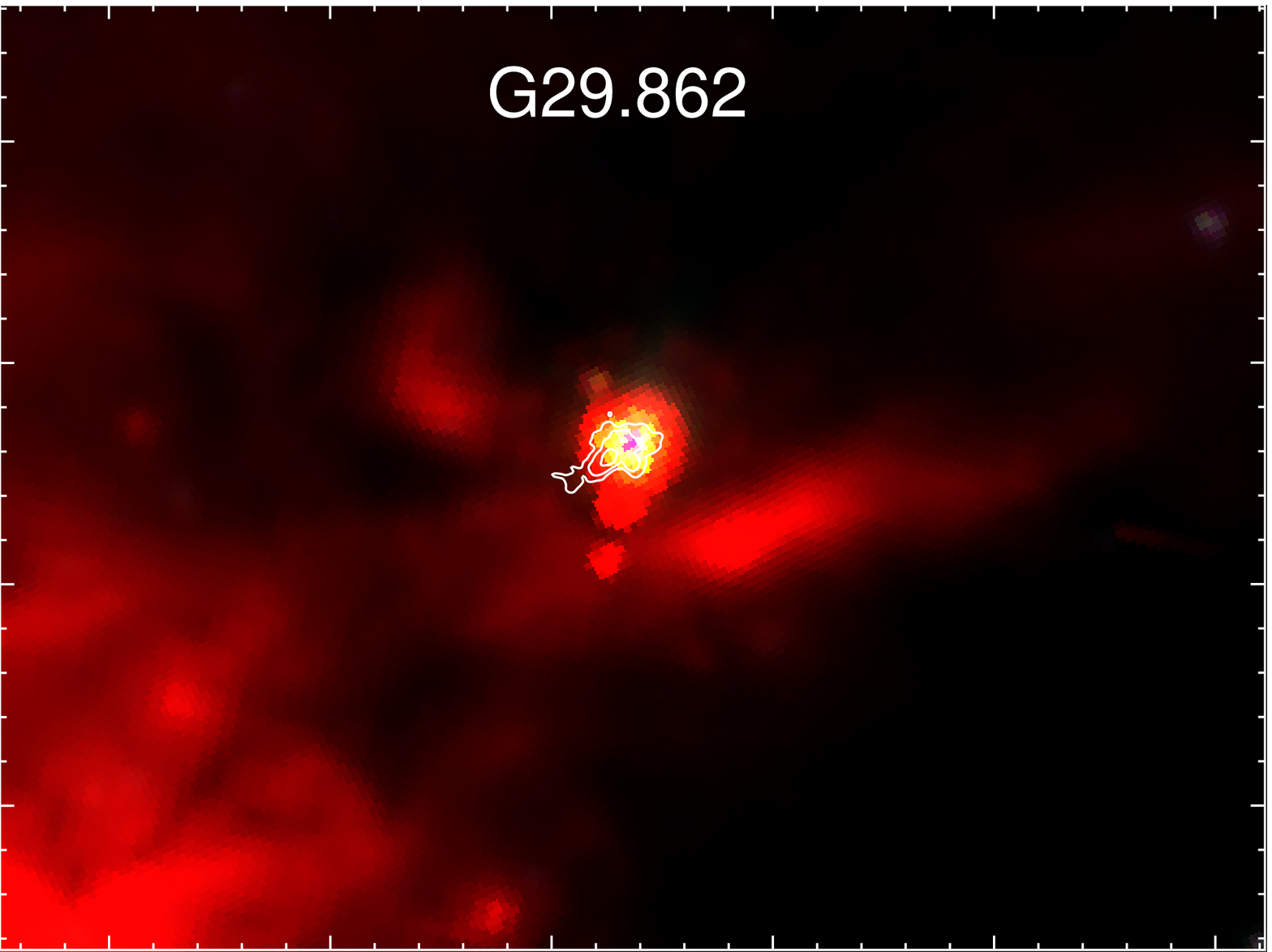}
    \includegraphics[width=4.3cm]{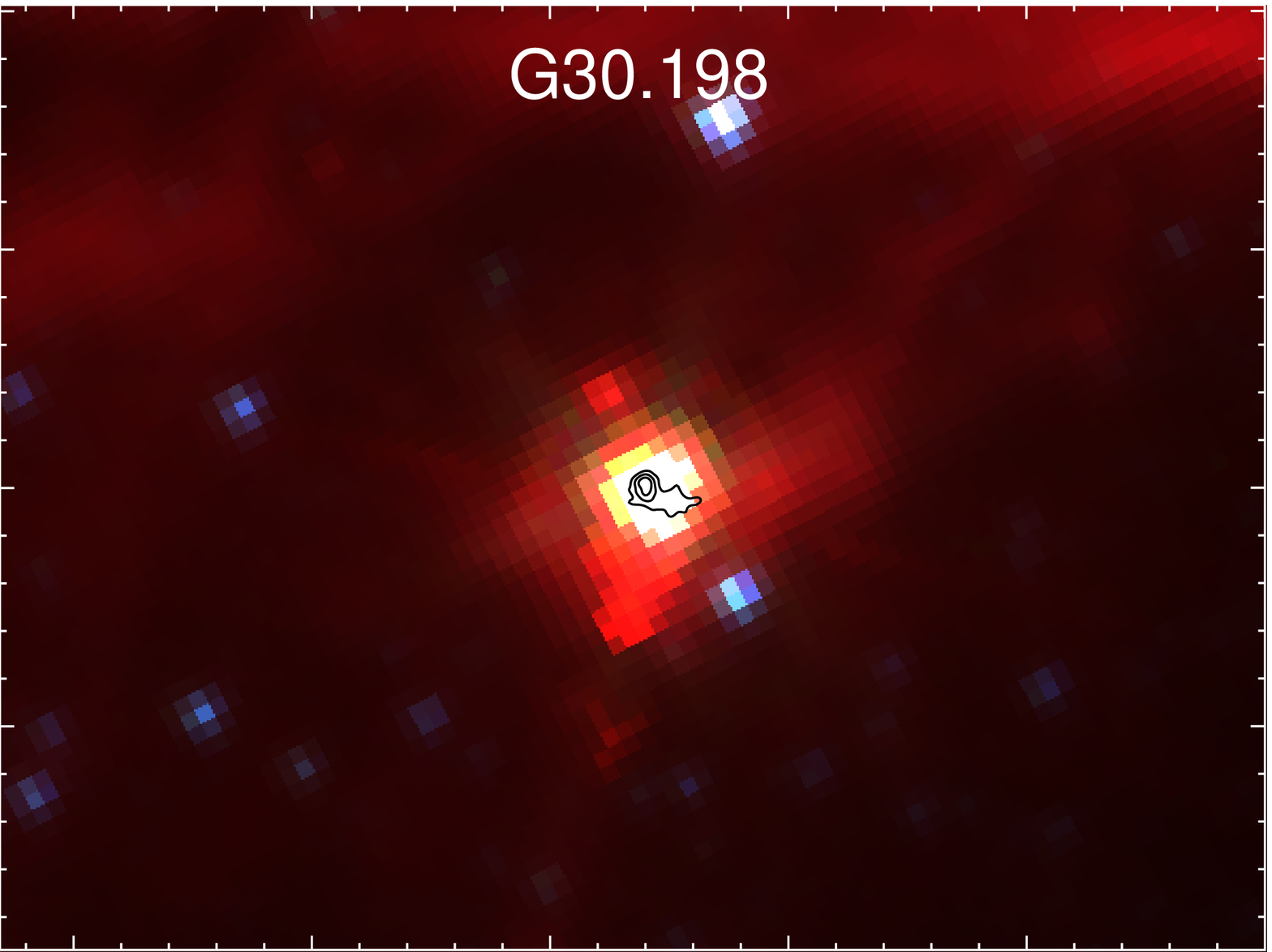}
    \includegraphics[width=4.3cm]{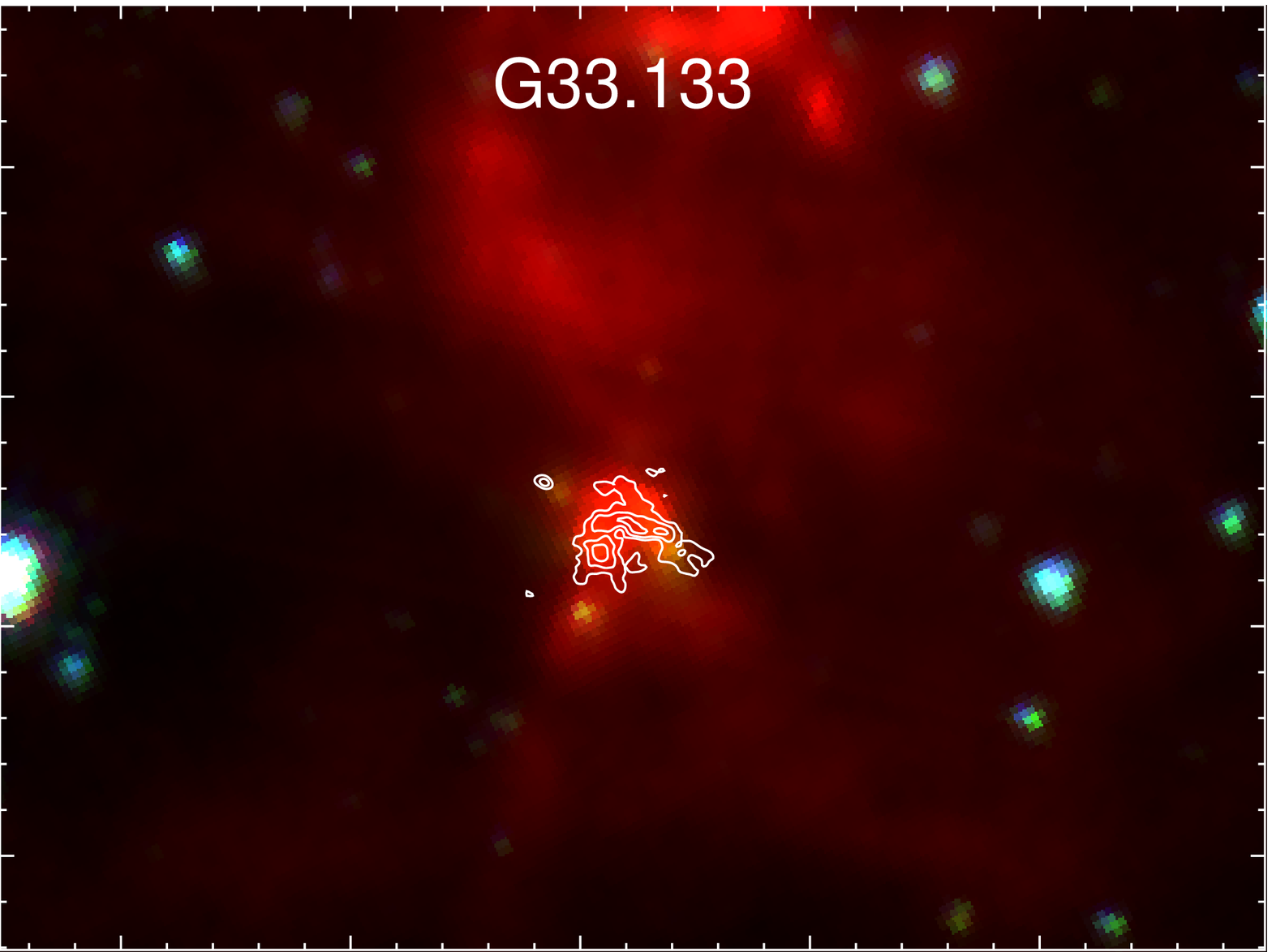}
    \includegraphics[width=4.3cm]{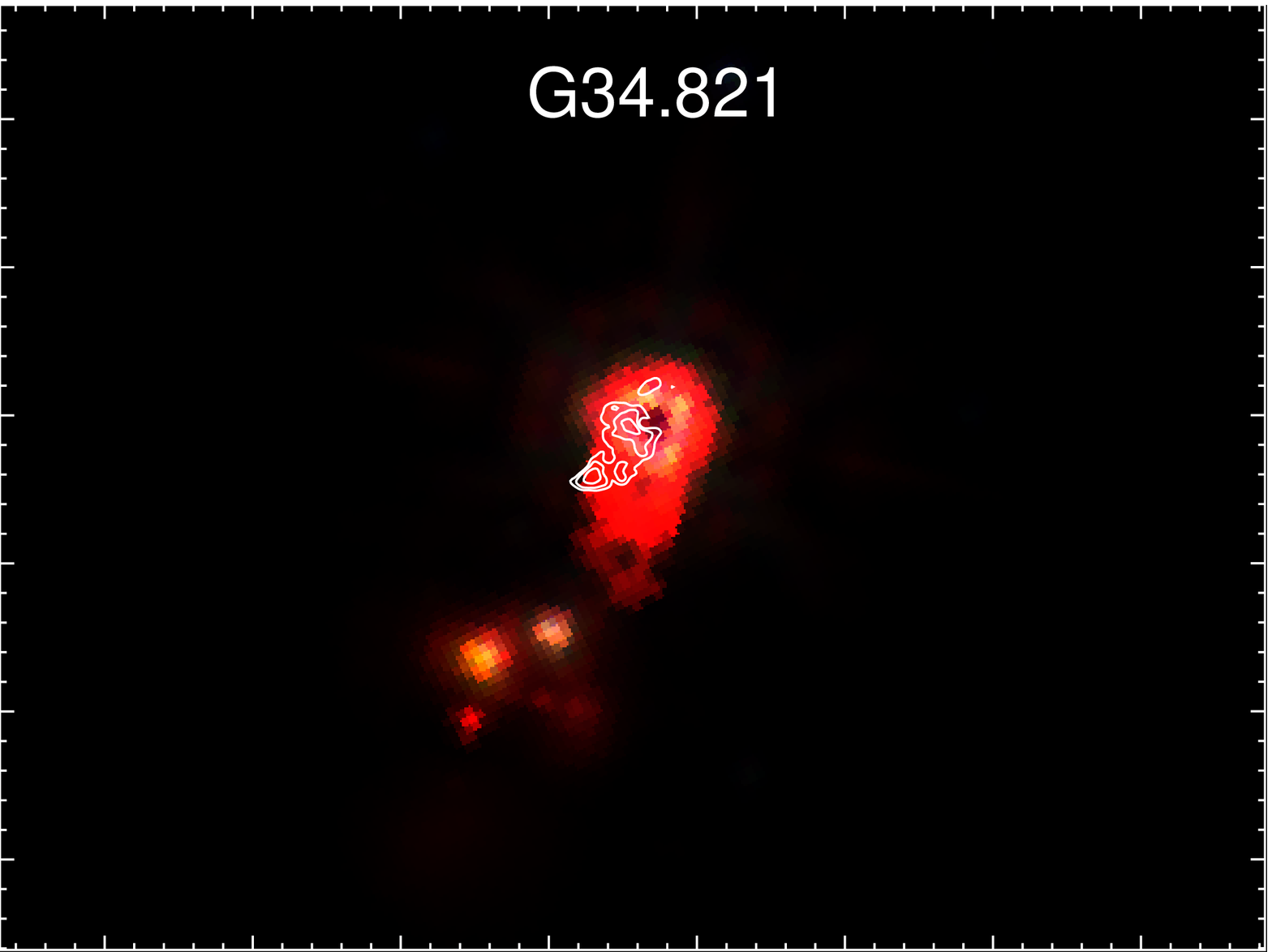}
    \caption{Three-colour image presenting the IR emission at 8 $\mu$m (red), 4.5 $\mu$m (green), and 3.6 $\mu$m (blue) obtained from {\it Spitzer}-IRAC. The contours represent the averaged CN emission as presented in Fig.\,\ref{cnIR}. The fields of view are about 1.5 arcmin for all cases but G23.389 that it is about 3 arcmin. }
\label{Spitzer}
\end{figure}

The near-IR emission usually is useful to identify the central point-like sources of 
the young stellar objects (YSOs) and traces diffuse emission arising mainly from warm dust and scattered light in the material surrounding the protostar(s), particularly in cavities carved out by jets on an infalling envelope of material (e.g. \citealt{muze13,bik06,bik05}). Even though the near-IR observations can sometimes miss deeply embedded protostars, 
the lack of near-IR emission or a point-like source at this wavelength in a molecular core can indicate that it is an starless
core, which represents the transition between a diffuse molecular cloud and the next generation of stars to form therein \citep{schnee12}. Thus, it is interesting to analyze the {\it Ks}-band emission in comparison with the ALMA data with the aim of finding any connection relating the extended CN and near-IR emissions, and near-IR point-like sources with the 1.3 mm continuum cores. This comparison is presented in Fig.\,\ref{cnIR} for each source. Additionally, in order to make a similar comparison, but in a larger spatial scale, in Fig.\,\ref{Spitzer} we present similar panels as presented in Fig.\,\ref{cnIR} but using the 3.6, 4.5, and 8 $\mu$m emission observed with {\it Spitzer}-IRAC. While in some cases the emission is saturated, this is still useful to analyze the sources and the surrounding medium.

\section{Discussion}

A first important result is that by analyzing the spatial morphology of the CN
emission at each source (see Fig.\,\ref{cnAve}), we can conclude that the CN 
traces both molecular condensations and diffuse and extended gas surrounding them. It is worth noting that, in general, these condensations traced by the maximums of the CN emission do not spatially coincide with the peaks of the continuum emission at 1.3 mm, that may have contributions from both ionized gas emission and from the dust \citep{hernandez14} usually associated with
the smallest structures related to star formation. This phenomenon was also observed towards IRAS 20293+3952 by \citet{beuther04} using interferometric data of the CN N=1--0 line with a similar angular resolution as the data presented in this work. The authors found that the CN emission avoids the core  peaks traced by the millimeter continuum emission, and based on
that optical depth effects appear to be negligible, they suggest two likely main
causes for the lack of CN emission towards the cores: (1) the source is too deeply embedded to excite the surface layers of the potential disk, and (2) the source is in such an early evolutionary stage that it does not generate enough UV photons to produce CN emission. Taking into account that some spectra presented in Fig.\,\ref{cnLinea} show probably absorption or self-absorption features, and considering the works of \citet{qiu12} and \citet{zapata08} in which the CN emission presents kinematical indications of infalling envelopes of gas, we point out that the non coincidence between the CN bulk emission and the cores traced by the continuum deserves a deeper analysis (see Sect.\,\ref{sptabs}). 

From Fig.\,\ref{cnIR}, it can be observed that in five sources (G20.762, G23.389, G29.862, G30.198, and G34.821) the continuum at 1.3 mm coincides with a point-like source (or almost point-like) at the near-IR, while in the others sources, there is not any near-IR emission associated with the 1.3 mm continuum. This suggests that our core sample is likely composed by regions at different stages of evolution in star formation: YSOs already emitting at the near-IR, and the `IR quiet' sources representing YSOs in an earlier stage of evolution, or even, starless cores (e.g. \citealt{motte18,schnee12}). 

The difference in the evolutionary stages of the sources can be also appreciated in Fig.\,\ref{Spitzer}. The IR emission obtained from {\it Spitzer}-IRAC shows that the above mention more evolved sources present point-like morphologies in these bands. Moreover, the excess in 8 $\mu$m suggests photodissociated gas due ultraviolet photons from young stars. In the case of G13.656 it is important to note that, as also shown in 
Fig.\,\ref{cnIR}, there is a point-like source in the field not related to the CN emission and the 1.3 mm continuum source.

It is important to remark that in all cases, extended CN emission appears. In the cases that we observe a point-like source at near-IR related to the 1.3 mm continuum, the related CN should be tracing gas in the near vicinity of the protostar as proposed by \citet{beuther04} and \citet{jor04}, 
and in the case of the `IR quiet' sources, the CN emission may arise from a molecular core in a pre-stellar phase. Indeed, it can be observed that in regions with near-IR point-like sources associated with the 1.3 mm continuum, the morphology of the CN emission appears, in general, more localized than in the `IR quite' sources. Whatever, we conclude that the presence of CN may be ubiquitous along the different star formation stages.

\subsection{Analysis of sources with spectral absorption features}
\label{sptabs}

The spectra obtained towards sources G13.656, G20.747, G24.464, and G33.133 present
some absorption features (see Fig.\,\ref{cnAve}), and hence we decided to analyze them in more detail. Firstly, we 
note that these sources are those that their continuum cores are not associated with near-IR sources 
(see Fig.\,\ref{cnIR}) suggesting that, as mention above, they can be cores in a pre-stellar phase. In that sense,
the absorption features could be due to high optical depth effects. 
The cores with IR sources may have lower densities than the `IR quiet' cores  because outflow activity
could have generated cavities in the molecular envelopes \citep{wheel12}, and the high optical depth effects affecting the CN emission, if there are any, are not so evident. 

In addition, the spectra may present absorption features like regular or inverse P-cygni profiles, indicating expanding gas or that a gaseous envelope is falling into an accreting object as observed in CN transitions by \citet{qiu12} and \citet{zapata08}. If the absorption features appear repeatedly along the observed frequency/velocity range, also it could be occurring absorption in the diffuse gas along the line of sight against a continuum source. \citet{godard10} studied 
this phenomenon towards star-forming regions using lines of HCO$^{+}$, HCN, HNC, and CN from single dish observations.    

As mention in Sect.\,2, we discarded that such absorption features are due to the imaging process. However,
it is important to take into account that they also could be due to instrumental effects.  The interferometric
emission not combined with single dish data can be affected by the missing flux coming from more extended spatial-scales that are filtered out by the interferometer that can not be recovered even
with the most accurate imaging process. This issue can produce not only missing flux but also such absorption features
(e.g. \citealt{rodon12,beuther04}). Hence, it is necessary to analyze this possibility, or at least to take it into account, before to conclude anything about the nature of the observed spectral absorptions. Thus, for each of the mentioned sources, we analyze the spectra at the positions of the continuum cores. 



\subsubsection{At core G13.656-mm1}

\begin{figure}[h]
    \centering
    \includegraphics[width=8cm]{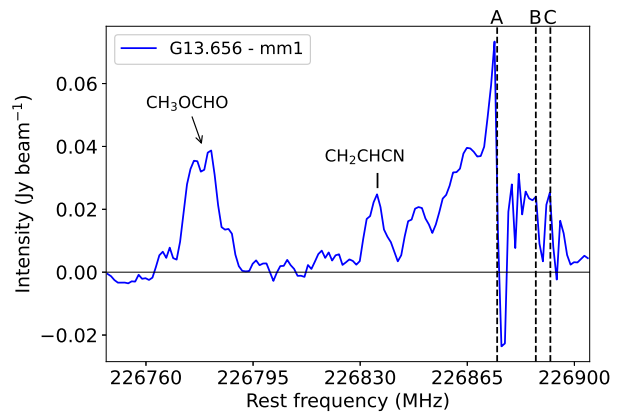}
    \caption{Spectrum obtained towards core G13.656-mm1. Vertical dashed lines named A, B, and C indicate the rest frequencies of the CN transitions (see Fig.\,\ref{cnLinea}). The emission of CH$_{3}$OCHO and CH$_{2}$CHCN is indicated.}
\label{mm13}
\end{figure}

Figure\,\ref{mm13} shows the spectrum obtained from a circle of 1\farcs5 in radius at the position of G13.656-mm1.
It can be observed a very complex spectrum in which a quite deep and narrow absorption feature related to CN appears at 226877 MHz. Additionally, it can be observed emission lines of CH$_3$OCHO (transitions 20(1,19)--19(1,18)E and A), and 
according to the NIST database, CH$_{2}$CHCN (transition $^{1}\nu_{11}$ 24(2,23)--23(2,22)), which are complex molecules usually detected towards hot 
molecular cores and pre-stellar cores \citep{jim16,herbst09,frie08}. The CH$_{2}$CHCN is a complex
cyanide whose observation and analysis, among other cyanides, can be used as a chemical clock of hot cores \citep{allen18}. Particularly, the CN is important for the formation of this molecular species, because it results from the reaction between CN and ethylene \citep{agundez08,herbst90}.

Even though the absorption feature at 226877 MHZ is narrow, it could be interpreted as a P-cygni profile suggesting expanding gas in this molecular core. However, we cannot discard that it could be due to the instrumental effect discussed above.

\subsubsection{At cores G20.747-mm1/2}

\begin{figure}[h]
    \centering
    \includegraphics[width=8cm]{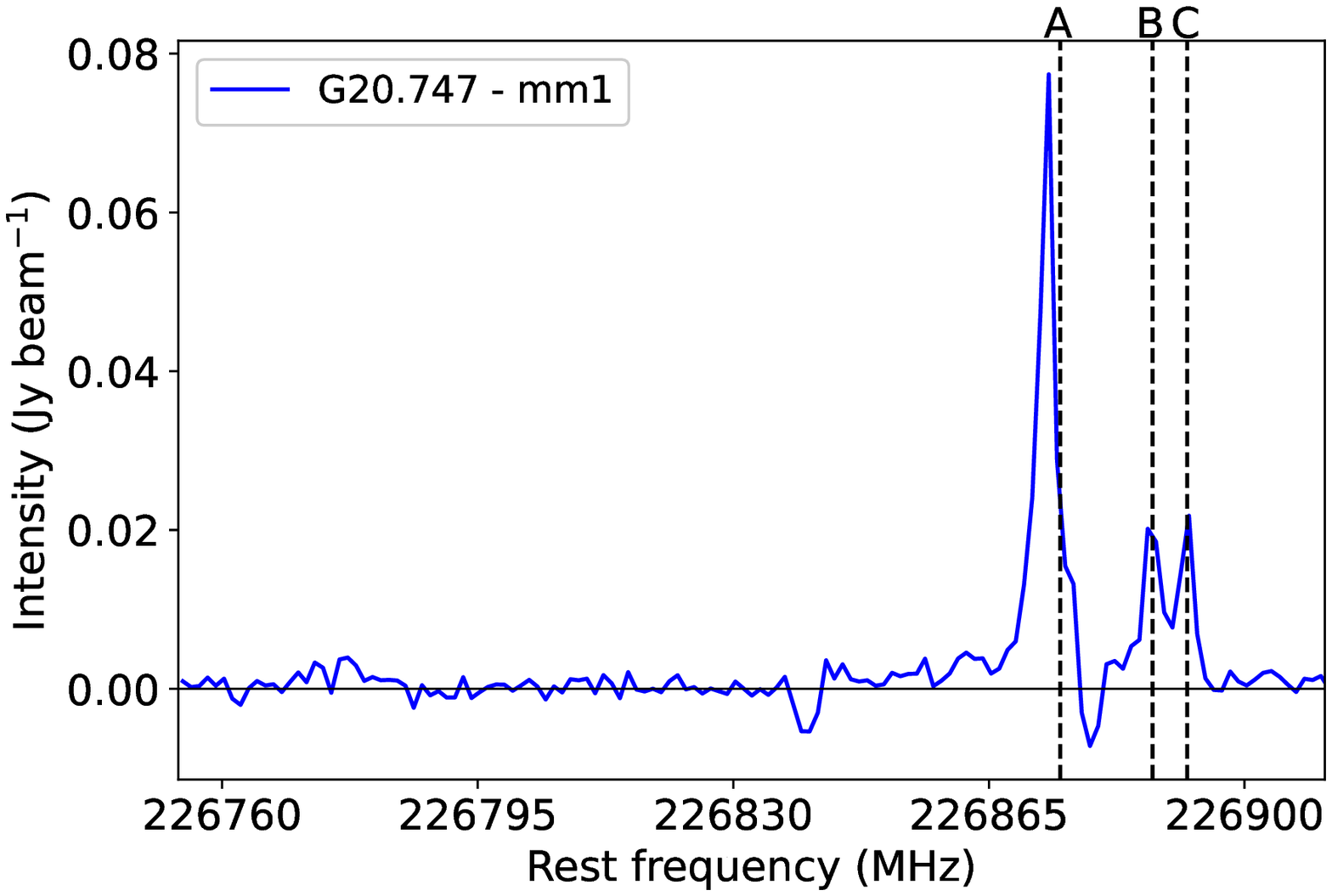}
    \includegraphics[width=8cm]{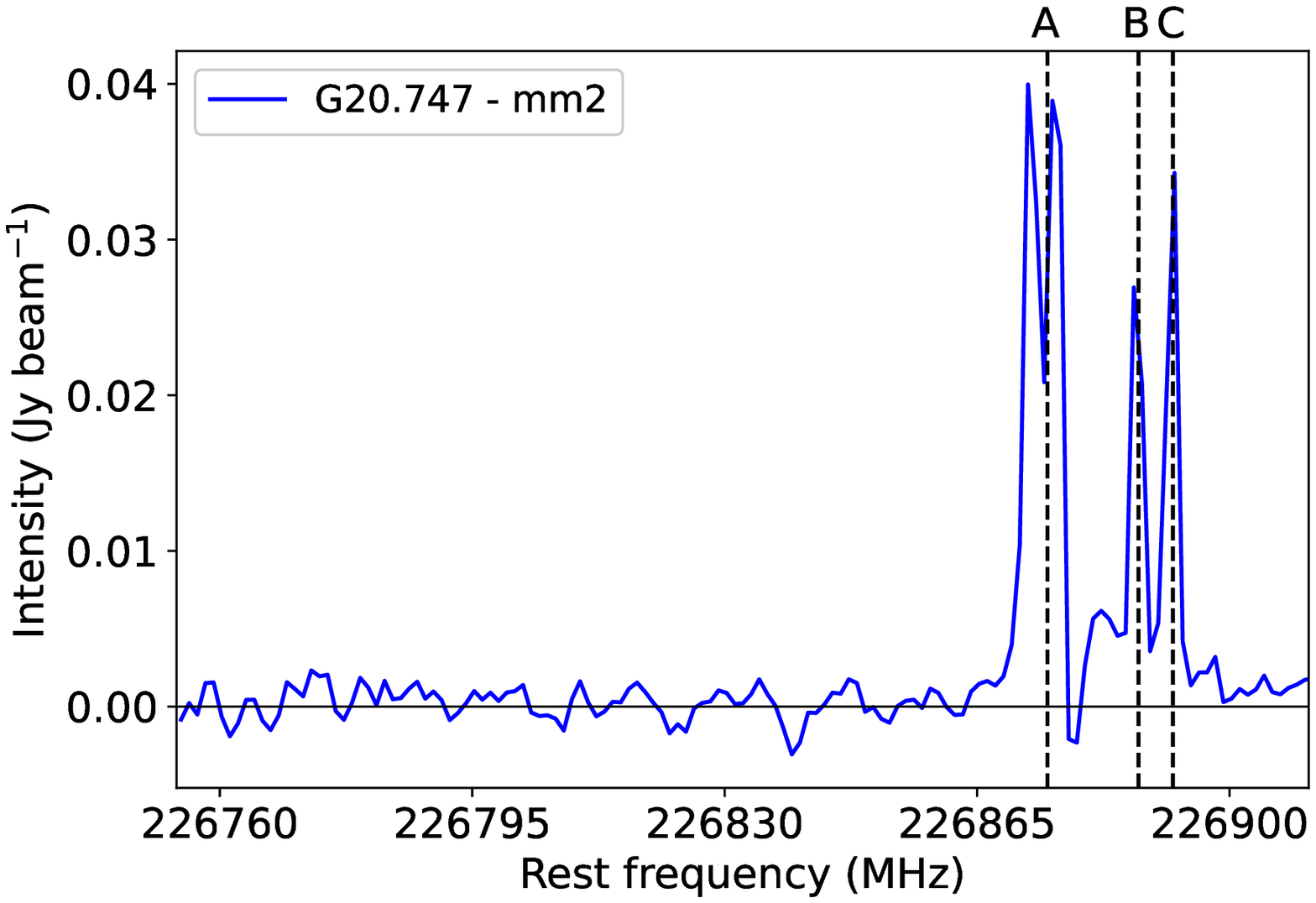}
    \caption{Spectra obtained towards core G20.747-mm1 and -mm2. Vertical dashed lines named A, B, and C indicate the rest frequencies of the CN transitions (see Fig.\,\ref{cnLinea}).}
\label{mm20}
\end{figure}

Figure\,\ref{mm20} displays the spectra obtained towards cores G20.747-mm1 and -mm2 from a circle of 1\farcs2 in radius. In both spectra, an absorption feature related to the CN main
component appears at 226877 MHz suggesting a P-cygni profile, and in the case of core G20.742-mm2, the CN main component has a dip at about 226873 MHz, which could be explained as a self-absorption suggesting that the emission is optically thick. As in the above case, we cannot discard that the P-cygni like feature is due to the mentioned instrumental effect.
But the dip at the CN main component coincides exactly with the systemic velocity of the source, suggesting that it 
could be due to a high optical depth.

The profile of G20.747-mm1 presents an small absorption at 226838 MHz, which does not appear in others regions,
showing that it is not due to an instrumental effect and it should be real. The frequency of this absorption is close to the that of the CH$_{2}$CHCN transition mentioned in the case of core G13.656-mm1, and hence, we wonder if it may be due to absorption in a molecular envelope containing this molecular species against the bright continuum.

\subsubsection{At core G24.464-mm1}

\begin{figure}[h]
    \centering
    \includegraphics[width=8cm]{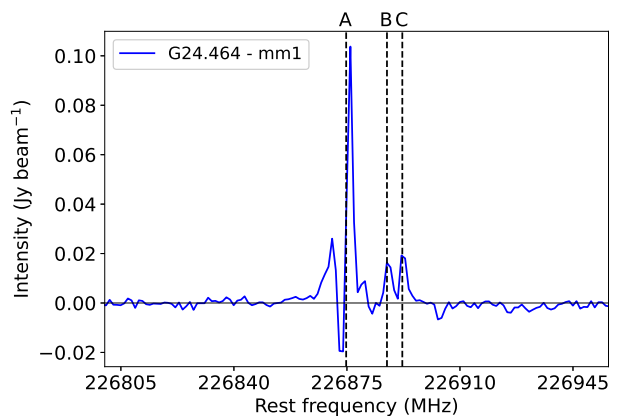}
    \caption{Spectrum obtained towards core G24.464-mm1. Vertical dashed lines named A, B, and C indicate the rest frequencies of the CN transitions (see Fig.\,\ref{cnLinea}).}
\label{mm24}
\end{figure}

The spectrum obtained from a circle of 1\farcs3 in radius at the position of core G24.464-mm1 is presented
in Fig.\,\ref{mm24}. It presents a conspicuous absorption dip at 226874 MHz, very close to the systemic velocity of 
the source, suggesting that it could be due to high optical depths, but missing flux coming from more extended spatial-scales filtered out by the interferometer cannot be discarded.

\subsubsection{At cores G33.133-mm1/2/3 }

\begin{figure}[h]
    \centering
    \includegraphics[width=8cm]{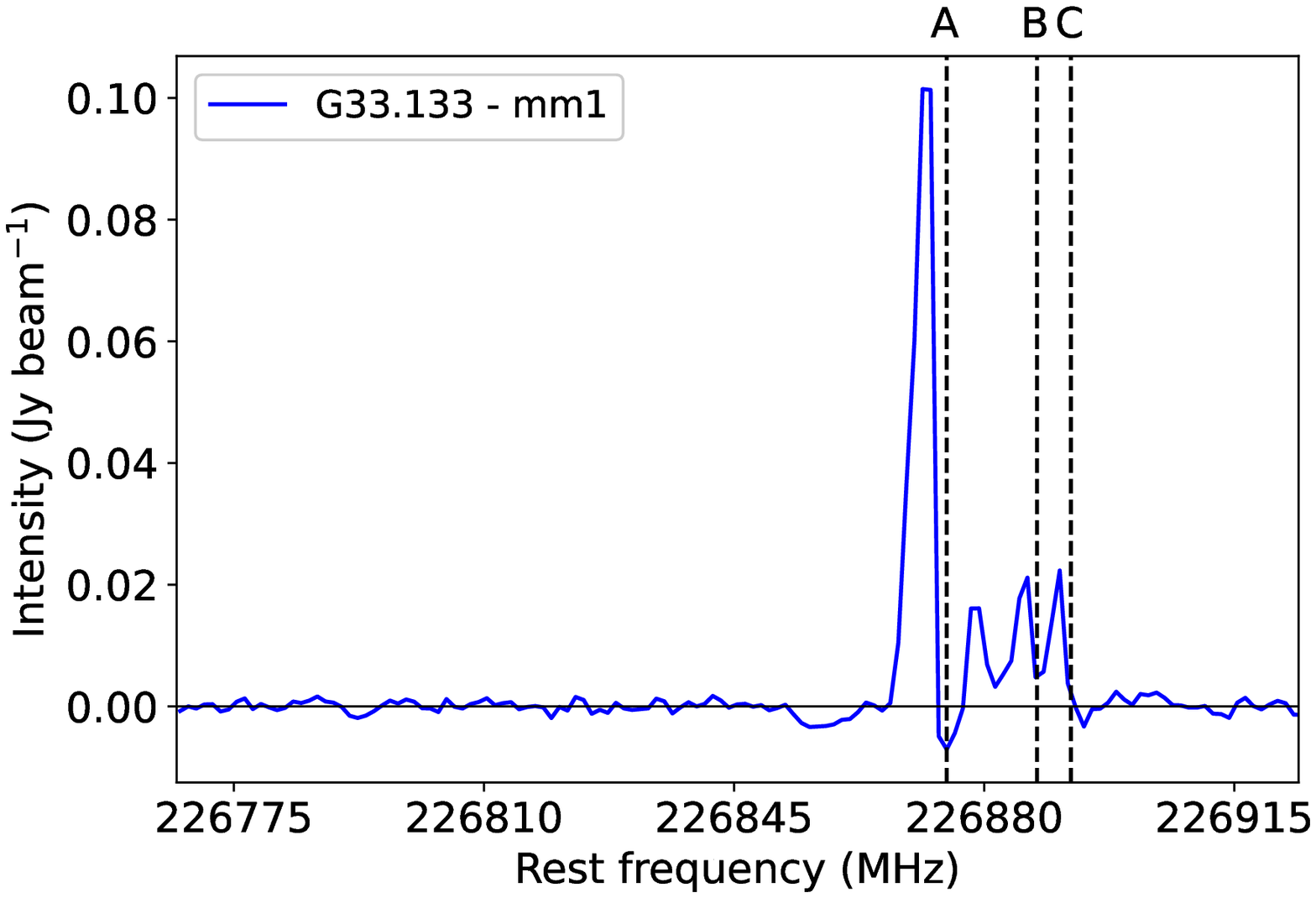}
    \includegraphics[width=8cm]{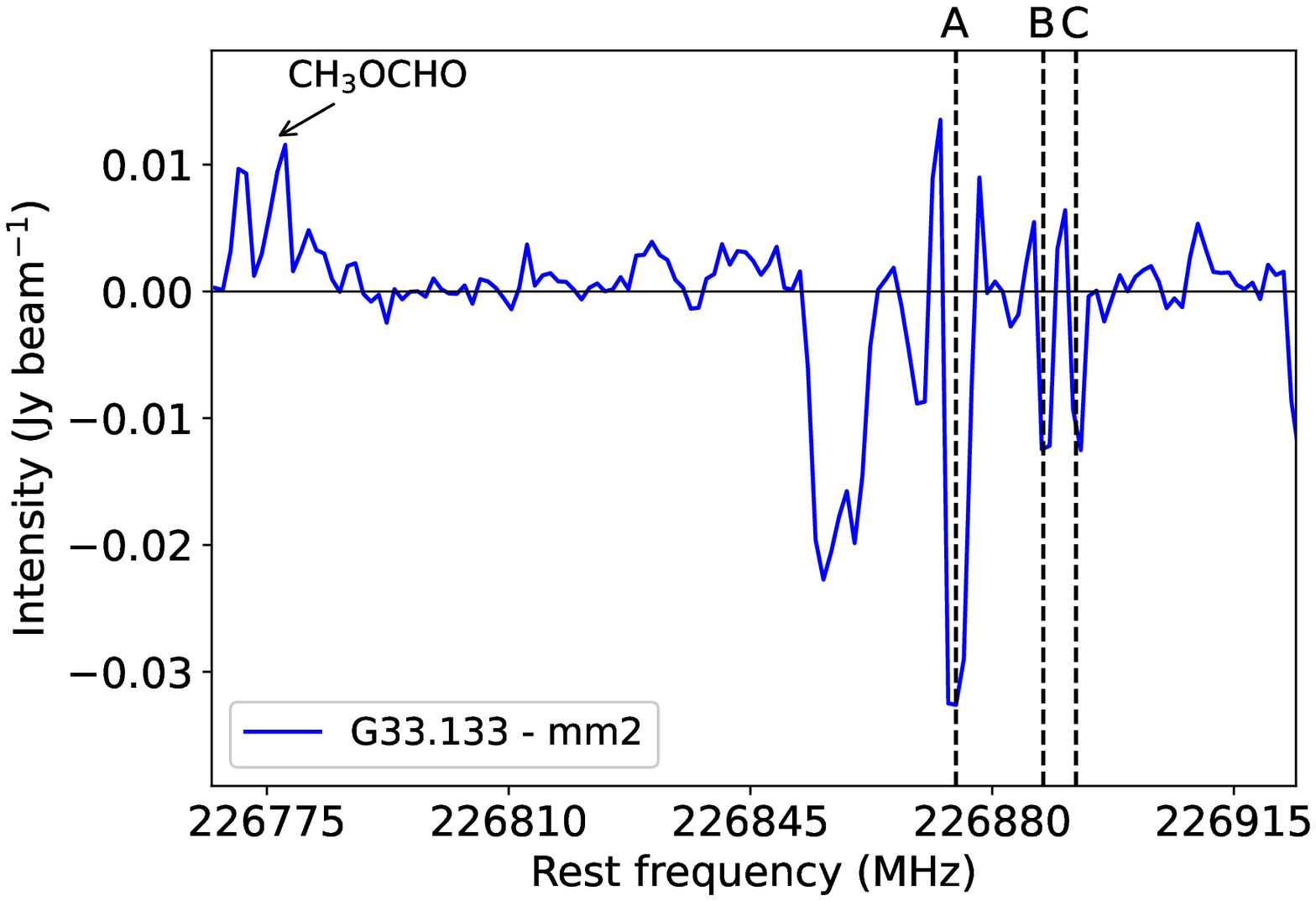}
    \includegraphics[width=8cm]{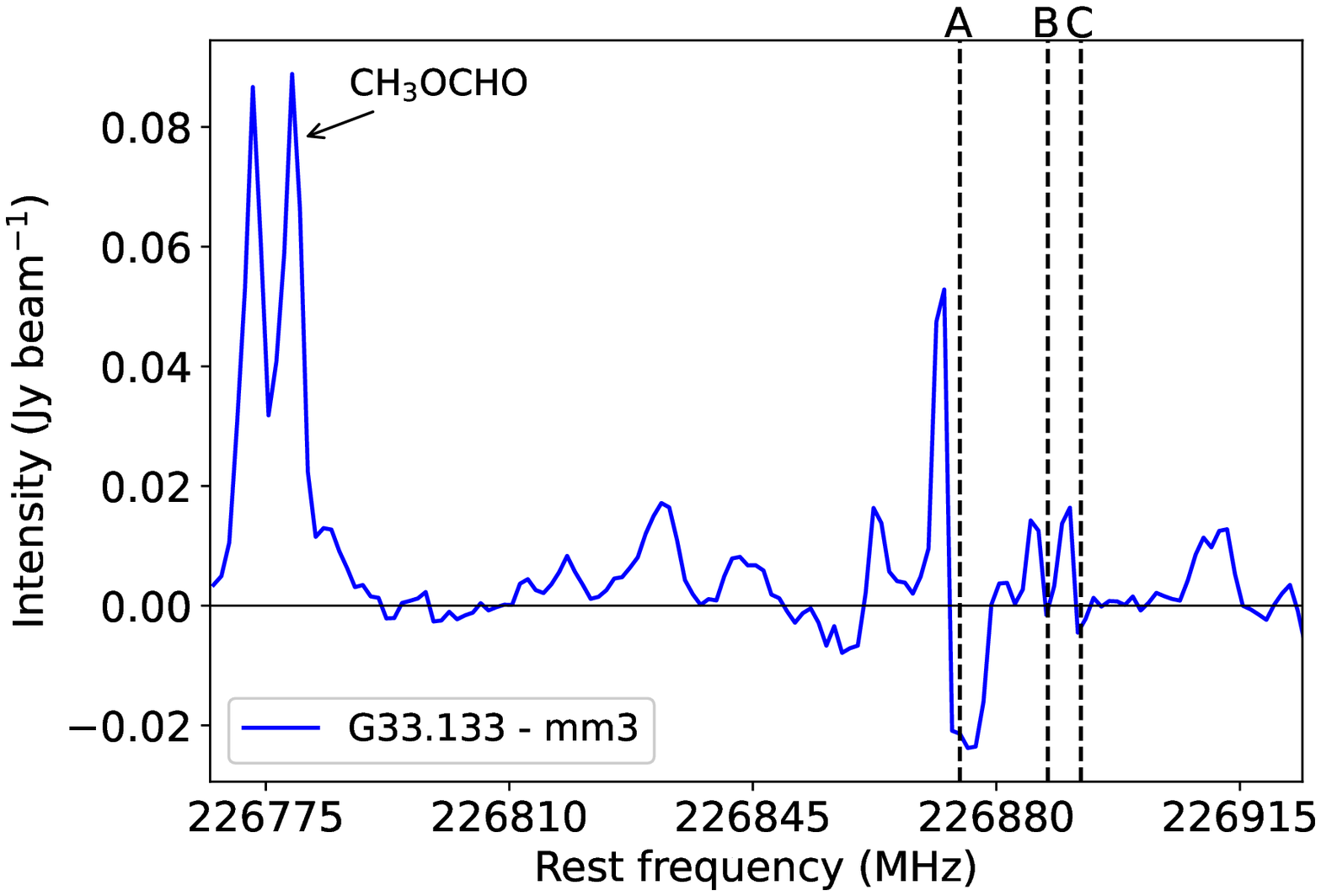}
    \caption{Spectra obtained towards cores G33.133-mm1,\,-mm2, and-mm3. Vertical dashed lines named A, B, and C indicate the rest frequencies of the CN transitions (see Fig.\,\ref{cnLinea}). The emission of CH$_{3}$OCHO is indicated.}
\label{mm33}
\end{figure}

Spectra obtained from a circle of 1\s~in radius towards cores G33.133-mm1, -mm2, and -mm3 are displayed in Fig.\,\ref{mm33}. All the spectra present absorption at the main CN line. In the cases of G133.133-mm2 and -mm3 it could be due to absorption in a molecular envelope against the bright continuum. Again, missing flux produced in the interferometric observation cannot be discarded in any of these three cores. Emission of CH$_{3}$OCHO is evident in G33.133-mm2 and -mm3. From the spectra obtained towards these cores, we can say that it is clear that the non-coincidence between CN maximums and the continuum cores can be due to a combination of physical absorption with the described instrumental issue. 

The spectrum of G33.133-mm2 presents a conspicuous absorption at 226856 MHz, which also appears in the spectrum of G33.133-mm1. This absorption does not seem to be produced by instrumental effects.
It should be real, and we wonder the same possibility as in the case of the absorption found in G20.747-mm1.

\bigskip

In conclusion, in most of the analyzed absorption features cannot be discarded that they are produced, totally, or 
in part, by the missing flux coming from more extended spatial-scales that are filtered out by the interferometer. 
In the case of the other sources, i.e. the sources that 
does not present any absorption feature in the spectra shown in Fig.\,\ref{cnLinea}, we inspect the emission at the 
position of the associated 1.3 mm continuum cores, and they do not present any important absorption feature. It is possible that as these sources could be more
evolved, given that they present near-IR point-like sources, the surrounding molecular envelopes at the more 
extended spatial-scales can be more diluted/evacuated due to the action of winds and outflows, and hence, the mentioned instrumental effect and/or the probable optical depth effects are not so notorious. 

Finally, we found that in cores with the largest integrated flux density at 1.3 mm continuum emission (cores G13.656-mm1, G24.464-mm1, and G33.133-mm2) the deepest absorption features in the whole field of view related to the CN emission positionally coincides with the peaks of the continuum emission. This suggests that high optical depth effects can
be combined with the discussed instrumental issue.

\section{Summary and concluding remarks} 
\label{concl}

Given that the CN radical is very reactive with molecules possessing C=C double and C$\equiv$C triple bonds, it is involved in the formation of complex molecules in dark clouds, and in particular, in those related to star-forming processes. Hence, the study of its emission at small spatial scales towards massive protostars is important.
Using high-angular resolution data we investigate the CN emission in a sample of ten massive young stellar objects located at the first Galactic quadrant, and the main results found are summarized as follows.

(a) All analyzed sources present emission of CN in the transitions N$=$2--1 J$=$5/2--3/2 F$=$7/2--5/2, F$=$3/2--3/2 and F$=$1/2--1/2. We observe that the CN traces both molecular condensations and diffuse and extended gas surrounding them.

(b) In general, the molecular condensations traced by the maximums of the CN emission do not spatially coincide with the peaks of the continuum emission at 1.3 mm as found in other works.

(c) Our sample of sources are divided in those that present near-IR emission associated with the continuum at 1.3 mm, and those that are `IR quiet' sources, suggesting that they are protostellar objects at different stages of evolution. The CN is present at both, suggesting that this radical 
may be ubiquitous along the different star formation stages, and hence it may be involved in different chemical reactions occurring along the time in the formation of the stars. For instance, the CN participates in the formation of
vinyl cyanide (CH$_{2}$CHCN), molecule detected in core G13.656-mm1.

Finally, it is worth noting that a molecular line emission analysis at such small spatial scales can be affected by 
the missing flux coming from more extended spatial-scales that are filtered out by the interferometer. This
is an important issue because it can introduce artificial absorption features in the spectra that can be misinterpreted.
We found that `IR-quite' sources are in general more affected by this issue, and we suggest that in the more evolved sources, the surrounding molecular envelopes at the more extended spatial-scales can be more diluted/evacuated due to the action of winds and outflows, and hence, this effect is not so notorious.  We cannot discard high optical depth effects combined with missing flux. Interferometric observations of the analyzed CN line in a more compact configuration would be useful to properly evaluate this issue.

\begin{acknowledgements}

We thank the anonymous referee for her/his very useful comments. 
A.M. and M.B.A. are doctoral fellows of CONICET, Argentina. 
S.P. and  M.O. are members of the Carrera del Investigador Cient\'\i fico of CONICET, Argentina. 
This work is based on the following ALMA data: 2015.1.01312.S. ALMA is a partnership of ESO (representing its member states), NSF (USA) and NINS (Japan), together with NRC (Canada), MOST and ASIAA (Taiwan), and KASI (Republic of Korea), in cooperation with the Republic of Chile. The Joint ALMA Observatory is operated by ESO, AUI/NRAO and NAOJ.

\end{acknowledgements}

%
%

\bibliographystyle{aa}  
\bibliography{ref}
\IfFileExists{\jobname.bbl}{}
{\typeout{}
\typeout{****************************************************}
\typeout{****************************************************}
\typeout{** Please run "bibtex \jobname" to optain}
\typeout{** the bibliography and then re-run LaTeX}
\typeout{** twice to fix the references!}
\typeout{****************************************************}
\typeout{****************************************************}
\typeout{}
}
\label{lastpage}

\end{document}